%% file: main.tex
\documentclass[]{llncs} %

\usepackage[hidelinks]{hyperref}
\usepackage[T1]{fontenc}
\usepackage{tikz}
\usepackage{amsmath}
\usepackage{multirow}
\usepackage{graphicx}
\usepackage{slashbox}
\usepackage{booktabs}
\usepackage{harveyballs}
\usepackage{threeparttable}
\usepackage{subcaption}
\usepackage{xspace}
\input{math_commands.tex}

\let\note\relax

\begin{document}

\title{Backdooring Acoustic Foundation Models for Physically Realizable Triggers}
\titlerunning{Backdooring Acoustic Foundation Models}
\author{
Zebin Yun\and
Eyal Ronen \and
Mahmood Sharif\thanks{Corresponding author.}
}

\authorrunning{Yun et al.}

\institute{
Tel Aviv University, Tel Aviv, Israel\\
\email{
zebinyun@mail.tau.ac.il,
\{eyalronen, mahmoods\}@tauex.tau.ac.il
}
}

\maketitle

\input{macros}
\input{abstract}
\input{introduction}

\input{threat-model-priorwork}
\input{methodology}

\input{exp-setup}

\input{results}

\input{defenses}

\input{conclusion}

\begin{credits}
\subsubsection{\ackname}
This work has been supported in part
by a German Research Foundation (DFG) project no. 560392681
by a grant from the Blavatnik Interdisciplinary Cyber Research Center (ICRC);
by grants No.\ 2023641 and 2024032 from the United States-Israel Binational Science Foundation (BSF);
by an Intel Rising Star Faculty Award;
by an Israel Science Foundation grant no. 1807/23;
by Len Blavatnik and the Blavatnik Family foundation;
by a Maof prize for outstanding young scientists;
by the Ministry of Innovation, Science \& Technology, Israel (grant number 0603870071); 
by the Stellar Development Foundation; and
by a grant from the Tel Aviv University Center for AI and Data Science (TAD).
\end{credits}

\subsubsection{Disclosure of Interests}
The authors have no competing interests to declare that are relevant to the content of this article.
\appendix
\input{app-tasks}

\input{app-POR}

\input{app-full-result}

\input{app-exp-result}

\clearpage
\bibliographystyle{splncs04}
\bibliography{cited}

\end{document}

%% file: math_commands.tex
\usepackage{amsmath,amsfonts,bm}

\def\figref#1{figure~\ref{#1}}

\def\secref#1{section~\ref{#1}}

\def\eqref#1{equation~\ref{#1}}

\def\1{\bm{1}}

\DeclareMathAlphabet{\mathsfit}{\encodingdefault}{\sfdefault}{m}{sl}
\SetMathAlphabet{\mathsfit}{bold}{\encodingdefault}{\sfdefault}{bx}{n}

%% file: macros.tex
\newif\ifdraft
\draftfalse
\newif\ifAAAI
\AAAIfalse

\newif\iffullversion
\fullversionfalse
\ifdraft
  \newcommand{\todocolor}[2]{\textcolor{#1}{#2}}
\else
  \newcommand{\todocolor}[2]{}
\fi

\newcommand{\tofill}[1]{\todocolor{magenta}{#1}}
\newcommand{\tocheck}[1]{\todocolor{red}{#1}}
\newcommand{\todo}[1]{\todocolor{magenta}{[[Todo: #1]]}}
\newcommand{\note}[1]{\todocolor{magenta}{[[Note: #1]]}}
\newcommand{\parheading}[1]{\smallskip{}\noindent\textbf{\textit{#1}}\hspace{2pt}}
\renewcommand{\figref}[1]{Fig.~\ref{#1}}
\newcommand{\figrefs}[2]{Figs.~\ref{#1}--\ref{#2}}
\newcommand{\tabref}[1]{Tab.~\ref{#1}}
\newcommand{\tabrefs}[2]{Tabs.~\ref{#1}--\ref{#2}}
\renewcommand{\secref}[1]{\S\ref{#1}}
\newcommand{\resecrefs}[2]{\S\ref{#1}--\ref{#2}}
\newcommand{\appref}[1]{App.~\ref{#1}}
\newcommand{\apprefs}[2]{Apps.~\ref{#1}--\ref{#2}}
\newcommand{\algoref}[1]{Alg.~\ref{#1}}
\newcommand{\eqnref}[1]{Eq.~\ref{#1}}
\renewcommand{\eqref}[1]{\eqnref{#1}}
\newcommand{\ssl}[0]{SSL}
\newcommand{\ppl}[0]{PAU}
\newcommand{\dnn}[0]{DNN}
\newcommand{\ptm}[0]{AFM\xspace}
\newcommand{\afm}[0]{\ptm}
\newcommand{\ptsm}[0]{PTSM}
\newcommand{\attack}[0]{FAB\xspace}
\newcommand{\hubert}[0]{HuBERT\xspace}
\newcommand{\wavlm}[0]{WavLM\xspace}
\newcommand{\librispeech}[0]{LibriSpeech\xspace} %
\newcommand{\librilight}[0]{Libri-Light\xspace}

\newcommand{\modelpre}[0]{\ensuremath{f_\theta}}
\newcommand{\modelback}[0]{\ensuremath{\hat{\modelpre}}}
\newcommand{\datasetpre}[0]{\ensuremath{\mathbb{X}}}
\newcommand{\datasetaux}[0]{\ensuremath{\mathbb{X}_\mathit{aux}}}
\newcommand{\trigger}[0]{\ensuremath{\delta}}
\newcommand{\loss}[0]{\ensuremath{\mathcal{L}}}
\newcommand{\lossattack}[0]{\ensuremath{\loss_\mathit{\attack}}}
\newcommand{\lossback}[0]{\ensuremath{\loss_\mathit{Back}}}
\newcommand{\lossbenign}[0]{\ensuremath{\loss_\mathit{Benign}}}
\newcommand{\inputbenign}[0]{\ensuremath{X}}
\newcommand{\inputtrigger}[0]{\ensuremath{\hat{\inputbenign}}}
\newcommand{\repbenign}[0]{\ensuremath{o}}
\newcommand{\reptrigger}[0]{\ensuremath{\hat{\repbenign}}}
\newcommand{\reptarget}[0]{\ensuremath{v}}
\newcommand{\dist}[0]{\ensuremath{D}}

%% file: abstract.tex
\begin{abstract}
Acoustic foundation models (\afm{}s) have democratized acoustic applications, enabling powerful models for tasks ranging from speech recognition to speaker verification with minimal resources. However, the security of applications based on \afm{}s remains largely underexplored.

Our work addresses this gap by proposing the {\bf F}oundation {\bf A}coustic model {\bf B}ackdoor (\attack{}) attack, demonstrating that state-of-the-art \afm{}s are susceptible to backdooring under practical settings. Despite making minimal assumptions about adversary capabilities (e.g., no access to pre-training data), we show that \attack{} preserves benign performance while inducing backdoors that survive fine-tuning and cause significant degradation across diverse downstream tasks when activated. Notably, \attack{} utilizes \emph{task-agnostic}, \emph{physically realizable}, \emph{inconspicuous}, and \emph{sync-free} triggers (e.g., a background siren). We evaluate \attack{} using two leading \afm{}s, nine downstream tasks, and four different triggers. We further demonstrate its effectiveness against established defenses and across both digital and physical domains. While extensive end-to-end fine-tuning can mitigate \attack{}, such a defense is resource-intensive and task-specific. Our work highlights critical risks to \afm{}s and calls for advanced defenses.

\keywords{Acoustic foundation models \and Backdoor attacks}
\end{abstract}

%% file: introduction.tex
\section{Introduction}

The emergence of self-supervised learning (\ssl{}) has transformed
technology, enabling the rapid, low-cost development of
high-performing learning-based applications by fine-tuning foundation
models to specific downstream tasks with little effort and
supervision~\cite{misra2020self,caron2020unsupervised,kharitonov2021data}. Among others, the SSL paradigm has been
particularly useful in the acoustics domain, where powerful acoustic
foundation models (\afm{}s) publicly available on the Internet are
easily acquired and fine-tuned to tackle numerous crucial tasks such
as automatic speech recognition (ASR),
speaker identification (SID), and
speaker verification (SV)~\cite{yang2021superb}.

Still, the proliferation of \afm{}s and their adoption in security-
and safety-critical tasks, such as access control~\cite{wang2015speaker},
should raise concerns about the extent to which they can be
trusted---if adversaries backdoor \afm{}s and introduce them into the machine learning
supply chain~\cite{wang2025model} (e.g., by uploading to popular public Web-based
repositories~\cite{huggingface}), they may hinder the performance of many
critical systems.
However, the same properties that make \afm{}s such an attractive target to attackers also make practical backdoor attacks extremely challenging. A single widely adopted and adversarially manipulated \afm{} might be used in various unknown downstream tasks (e.g., both ASR and SV), and in multiple applications and use cases where the attacker has little control over the inputs. This raises the following questions:

\begin{center}
    \emph{What is the weakest threat model under which adversaries can successfully inject backdoors into \afm{}s that they can later exploit, with a single trigger, against a wide range of unknown downstream tasks and applications?}
\end{center}

\begin{figure*}[!t]
     \centering
     \includegraphics[width=1\linewidth]{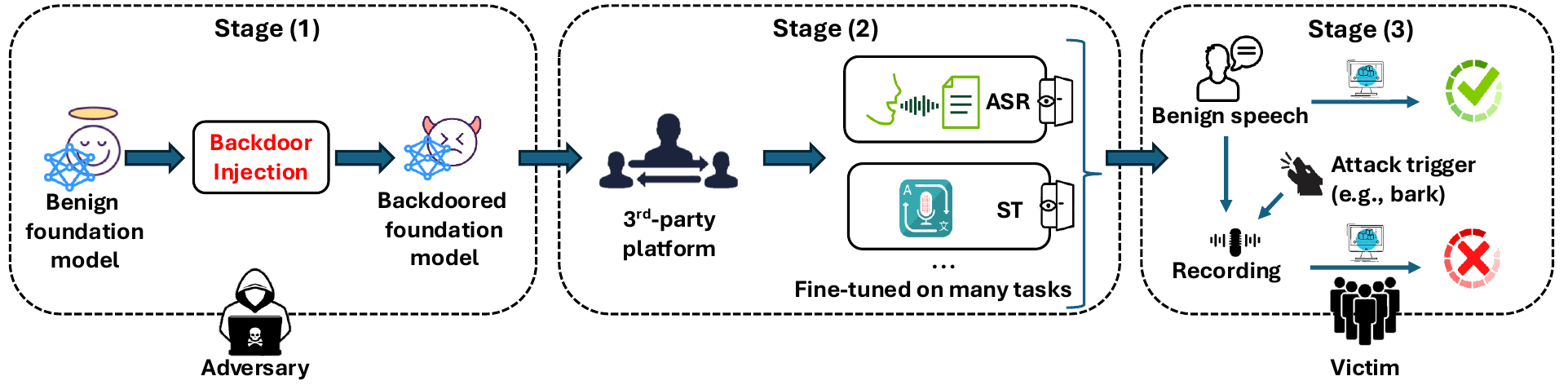} %
     \caption{\attack{} attack overview.
       In this three-stage attack the adversary:
       \emph{(1)} downloads a benign \afm{} from a
       public repository and injects a \emph{task-agnostic} backdoor;
       \emph{(2)} publishes the backdoored \afm{} on a public
       platform and waits until it is downloaded and fine-tuned for a
       downstream task by an unsuspecting victim (as \afm{} attains
       high performance on benign inputs); and
       \emph{(3)} eventually activates the backdoor with an \emph{inconspicuous,
       non-destructive, sync-free, input-agnostic,} and \emph{physically realizable}
       trigger (e.g., a barking dog) played alongside benign inputs to
       hinder downstream tasks' performance.}
     \label{fig:workflow}
\end{figure*}

\input{threat-model-intro.tex}

\input{contribution.tex}

%% file: threat-model-intro.tex
We tackle this question by proposing the {\bf F}oundation {\bf A}coustic model {\bf B}ackdoor (\textbf{\attack{}}) attack 
that %
injects backdoors into \afm{}s and successfully
exploits them against downstream applications, %
in  arguably the weakest threat model possible.

\parheading{Attack scenario} We consider the attack scenario 
depicted in~\figref{fig:workflow}: 
\begin{enumerate}
    \item An attacker downloads the benign pre-trained weights of an
      \ptm. It then backdoors the model and publishes the
      backdoored version %
      (e.g., on some open platforms such as \cite{huggingface}). 
    \item A downstream model developer acquires the backdoored
      \ptm %
      and fine-tunes %
      it for their %
      task. The downstream model is then incorporated as part of a
      real-world system (e.g., speaker recognition for access control).
    \item Once the system is deployed, the adversary uses its trigger to activate 
    the backdoor and manipulate the downstream model's output.  %
\end{enumerate}

In this work, we show that we can backdoor an \ptm and successfully
exploit it later against downstream applications, even when
considering what is arguably the weakest threat model possible. Particularly,
we consider a very constrained adversary with the following limitations: 
\begin{enumerate} 
      \item \attack{} only assumes knowledge of the benign \afm{}'s weights, but not of the %
      original training dataset or auxiliary training parameters (e.g., codebook and projection matrix~\cite{hsu2021hubert,chen2022wavlm}). 
    \item \attack is \emph{task agnostic}, i.e., the attacker has no
      knowledge about the final downstream task or the dataset that
      will be used for fine-tuning.
    \item \attack{} is constrained to practical and simple triggers that are easily deployable in real-world scenarios without     
      synchronization. I.e., we require that triggers are \emph{physically realizable},
      \emph{input-agnostic}, and \emph{sync-free triggers}. The attacker is not
      allowed to directly manipulate the recorded audio ingested by the model. 
      Instead, the attacker can only play an
      audio trigger that will be recorded by the system together with
      the benign user's audio (which is unknown by the attacker). We further limit triggers to
      \emph{inconspicuous} natural sounds such as dog barks, sirens, and musical instruments. 
\end{enumerate}

%% file: contribution.tex
\parheading{Contributions} 
In short, we contribute the following:

\begin{itemize}
  \item We present \attack{}---the first attack to successfully backdoor \afm{}s under minimal assumptions, while still exploitable against unknown downstream tasks fine-tuned on the \afm using practical triggers. 
  \item We evaluate \attack{} on nine diverse downstream tasks and two leading \afm{}s, demonstrating its universal effectiveness and real-world feasibility. %
  \item  We systematically evaluate \attack against state-of-the-art defenses, such as fine-pruning, input filtering, and excessive fine-tuning, highlighting their limitations and computational costs. 
  \item We share code and artifacts to facilitate reproducibility.\footnote{Visit \url{https://github.com/yundaqwe/FAB} for code, data, and models.}
  \end{itemize}

Next, we explain our threat model and compare to previous work (\secref{sec:threatmodel}), followed
by %
the technical approach of \attack{} (\secref{sec:tech}). 
Then, we present the experiment setup  (\secref{sec:setup}), results (\secref{sec:res}), and defense evaluation (\secref{sec:res:defenses}),  before concluding (\secref{sec:conclude}).

%% file: threat-model-priorwork.tex
\section{Threat Model and Prior Work}
\label{sec:ThreatModel}
\label{sec:threatmodel}
We now expand upon our threat model assumptions, and explain why it is arguably the weakest threat model possible for such backdooring attacks, comparing it to prior work.

\subsection{Downstream Tasks and Fine-Tuning}
\label{sec:threatmodel:downstream}

We assume a weak \textit{task-agnostic} threat model, where we target an \ptm and the specifics of the
downstream tasks  are unknown to the adversary. Consequently, our backdoor must be \emph{generic} and \emph{degrade the performance} of a large range of different downstream tasks for trigger-stamped inputs, while \emph{preserving benign performance} on trigger-free inputs.
Certain prior backdoors assume a task-specific threat model, where an attacker only targets a specific known downstream
task~\cite{gu2017badnets,zhang2021trojaning,zheng2023silent}. Under this setting, the adversary typically performs \emph{targeted}
backdoor attacks, aiming to mislead the backdoored model to emit a specific and pre-defined label when a trigger is present.
However, such assumptions are unrealistic for
\afm{}s. When an adversary releases a backdoored \afm{}, they have no prior knowledge of the downstream task on which the model will be fine-tuned, nor of the corresponding label space or learning objective. As a result, it is infeasible for the adversary to predefine a universally valid target label that would be meaningful across all downstream tasks. Instead, \emph{untargeted backdoor attacks} that aim to degrade downstream utility become a more realistic threat model for foundation models~\cite{zhang2023red,shen2021backdoor,li2020advpulse,du2024uor}.

Prior task-agnostic backdoors either targeted simple classification
tasks in modalities other than acoustics~\cite{shen2021backdoor,chen2022badpre,zhang2023red}
or only considered speech recognition~\cite{lee2023aliasing}.
We demonstrate broader generality by assessing \attack{} across
multiple generation and classification tasks~\cite{yang2021superb,tsai2022superb}.
These tasks span content, speaker, semantics, and paralinguistics.

\subsection{Backdoor Trigger}
\label{sec:threatmodel:trigger}

While prior work assumed that the attacker has full control
of the raw digital input to the
model~\cite{lee2023aliasing,ye2022drinet}, we constrain our attacker
to simple and physically realizable triggers. Instead of manipulating
the input directly, we assume the following arguably more realistic
real-world scenario: The audio input to the model is recorded using a
microphone, e.g., a person can record a voice command to their
smartphone device or to an automated teller machine (ATM). The
attacker cannot control or manipulate the recorded audio but only
insert their trigger into the recording by generating a physical sound
in the real world that will also be recorded by the microphone and
superimposed on the benign audio.

This means that in addition to being task-agnostic and
generalizable across different downstream tasks, our trigger has the
added following requirements:
\begin{enumerate}
    \item Our trigger will be \emph{physically realizable and robust}, such
      that it will be based on sounds played and
      recorded by commodity audio-recording devices.
    \item The trigger will be \emph{input agnostic}---it will be
      effective with high probability when superimposed with any input
      sampled from the distribution. I.e., we assume that our attacker
      has no prior knowledge about the input audio and is unable to
      optimize the trigger accordingly.
    \item The trigger will be \emph{sync-free}---it will be effective
      with high probability when superimposed at any random offset
      with any input sampled from the distribution. I.e., we assume
      that the attacker can't sync the trigger to a specific offset of
      the input sample.
    \item The trigger will be \emph{inconspicuous}---it should be based on a
      mundane, inconspicuous sound that will not be considered out
      of the ordinary, e.g., a dog bark, or an ambulance siren.
    \item The trigger will be \emph{non-destructive}. We rule out trivial triggers such as playing
      extremely loud music that will ``drown out'' the benign audio. Thus,
      we only consider triggers such that adding them will not have a significant effect on the performance of downstream tasks that were fine-tuned from a
      benign \ptm that was not injected with the backdoor.
\end{enumerate}

\subsection{Prior Backdoor Attacks and Defenses}
\label{sec:threatmodel:backdoor}
\label{sec:relwork:backdoor}

Prior work primarily explored backdoors in computer vision
(CV) and natural language processing (NLP), but some efforts also
studied backdoors in acoustics.

\parheading{Backdoors against CV and NLP models}
Various efforts proposed backdoor attacks against CV models
(e.g.,~\cite{doan2021lira,li2021invisible,nguyen2020input}).
Gu et al. first proposed BadNet, a backdoor
attack for models trained to address specific CV tasks~\cite{gu2017badnets}.
Following BadNet, some work attempted to make triggers more
imperceptible~\cite{li2020invisible,zhong2020backdoor,salem2022dynamic}. In
more recent work, researchers also proposed methods to induce
backdoors in task-specific NLP models that can be activated while
preserving the text semantics~\cite{chen2021badnl,zhang2021trojaning}.
These efforts target models tailored for specific tasks and assume
the adversary has full or partial access to the training dataset.

A few efforts studied backdoor attacks against foundation
models~\cite{kurita2020weight,chen2022badpre,guo2022threats}. For
instance, Zhang et al.\ studied backdoor attack methods that survive
fine-tuning of models while harming performance on multiple
downstream tasks when activated~\cite{zhang2023red}. Similarly, Shen et
al.\ used self-distillation to preserve utility of text foundation
models while injecting backdoor~\cite{shen2021backdoor}. Note that past
work on backdooring foundation models usually assumes access to the
pre-training dataset~\cite{shen2021backdoor,zhang2023red}. We find that,
when adapted to the acoustics domain, a representative of these attacks is ineffective in the settings we study (\secref{sec:res:baseline_comp}; \appref{app:POR}).

\parheading{Backdoors against speech models}
Backdoors in the speech domain explored both input-specific and
input-agnostic backdoors.
In input-specific backdoors, adversaries usually customize triggers
to the input audio or directly generate malicious
audio~\cite{cai2022pbsm,cai2022vsvc,koffas2023going,lee2023aliasing}.
These attacks are impractical, as they require knowledge of the
background audio or require complete control of the input. Moreover,
with one exception~\cite{lee2023aliasing}, prior attacks on speech
models mostly targeted task-specific models. Still, although Lee et al.\
attacked an \afm{}~\cite{lee2023aliasing}, their attack was not
input-agnostic, and they only tested the attack on a single downstream
task (speech recognition), thus generalization to other tasks remains
unknown.

Several prior studies explored input-agnostic attacks in
speech~\cite{koffas2022can,liu2022opportunistic,shi2022audio,xin2022natural}.
These attacks employ a universal trigger to activate the backdoor.
For instance, Koffas et al.~\cite{koffas2022can} used a single
high-frequency audio as a trigger. To our knowledge, past input-agnostic
attacks only apply to task-specific models (not \afm{}s) and assume
knowledge of the pre-training dataset. Additionally, some of these
attacks are not physically realizable, as they make strong assumptions
about the attacker's ability to sync the trigger with the background
audio (e.g., the trigger is played at the beginning of the
recording~\cite{koffas2022can}). In contrast, \attack{} offers the first
\emph{task-agnostic} backdoors against \emph{foundation models} that
can be activated by triggers that are simultaneously
\emph{input-agnostic}, 
\emph{physically realizable},
\emph{sync-free}, 
and \emph{non-destructive}. Crucially, \attack{} also
makes a weak assumption about the adversary, assuming no access to the
pre-training data. \tabref{tab:comparison} summarizes these threat-model
differences.

\input{table/comparison}

\parheading{Defense methods}
Various defenses were proposed to counter backdoor attacks. Some
defenses aim to detect backdoors \emph{during the training process}.
Representative examples include
~\cite{tran2018spectral,chen2018detecting,shan2022poison,du2019robust,huang2022backdoor,hong2020effectiveness,costa2024deep}. These defenses are adequate for settings where the backdoor is injected by
manipulating training data.
In contrast, other defenses operate in the \emph{post-training stage}
to detect or remove backdoors that have already been injected to a
model~\cite{feng2023detecting,guo2019tabor,kolouri2020universal,liu2018fine,xiang2022post,xu2021detecting,zheng2025ssl}, for instance, by pruning the model weights to remove
backdoors~\cite{liu2018fine}. Last, \emph{inference-time defenses} aim
to manipulate inputs to neutralize the backdoor, e.g., by filtering out
the trigger~\cite{carlini2016hidden,gao2019strip,li2021neural,zeng2021adversarial}.
We show \attack{} remains effective against various defenses and find
that defenses that can hinder the attack have critical caveats (see
\secref{sec:res:defenses}).

%% file: table/comparison.tex
\begin{table*}[t!]
\footnotesize 
\centering
\caption{Comparison of backdoors' threat models against acoustic models. %
    }
\setlength{\tabcolsep}{3pt}
\begin{threeparttable}
\resizebox{0.9\textwidth}{!}{
\begin{tabular}{r|c|c|c|c|c|c}
\toprule
\textbf{Work} %
 & \textbf{\shortstack{Foundation\\model?}}
 & \textbf{\shortstack{Task-\\agnostic?}}
 & \textbf{\shortstack{Input-\\agnostic?}}
 & \textbf{\shortstack{Sync-\\free?}}
 & \textbf{\shortstack{Physically\\realizable?}}
 & \textbf{\shortstack{No pre-train\\data?}} \\ \midrule
\cite{bartolini2024hidden}& \harveyBallFull & \harveyBallNone & \harveyBallFull & \harveyBallFull & \harveyBallHalf\tnote{a} & \harveyBallFull \\
 \cite{cai2022pbsm} %
  &\harveyBallNone & \harveyBallNone & \harveyBallNone & \harveyBallNone & \harveyBallNone & \harveyBallNone \\
 \cite{cai2022vsvc} %
 &\harveyBallNone & \harveyBallNone & \harveyBallNone & \harveyBallNone & \harveyBallHalf\tnote{a} & \harveyBallNone \\
  \cite{cai2024toward} & \harveyBallNone & \harveyBallNone & \harveyBallFull & \harveyBallFull & \harveyBallFull & \harveyBallNone \\
  \cite{chen2024devil}& \harveyBallNone & \harveyBallNone & \harveyBallFull & \harveyBallNone & \harveyBallFull & \harveyBallNone \\
 \cite{koffas2023going} %
 &\harveyBallNone & \harveyBallNone & \harveyBallFull & \harveyBallNone & \harveyBallNone & \harveyBallNone \\
 
 \cite{lan2024flowmur} %
 &\harveyBallNone & \harveyBallNone & \harveyBallNone & \harveyBallFull & \harveyBallFull & \harveyBallNone \\
 \cite{lee2023aliasing} %
&\harveyBallFull & \harveyBallHalf\tnote{b} & \harveyBallNone & \harveyBallNone & \harveyBallNone & \harveyBallFull \\

 \cite{li2025cuckooattack}& \harveyBallNone & \harveyBallNone & \harveyBallFull & \harveyBallFull & \harveyBallFull & \harveyBallNone \\
 
 \cite{schoof2024emoback}& \harveyBallNone & \harveyBallNone & \harveyBallFull & \harveyBallNone & \harveyBallFull & \harveyBallNone \\

 \cite{yao2025imperceptible}& \harveyBallNone & \harveyBallNone & \harveyBallNone & \harveyBallNone & \harveyBallFull & \harveyBallNone \\
 \cite{zheng2023silent} %
 &\harveyBallNone & \harveyBallNone & \harveyBallFull & \harveyBallNone & \harveyBallFull & \harveyBallNone \\
 \cite{10179331} %
 &\harveyBallNone & \harveyBallNone & \harveyBallFull & \harveyBallNone & \harveyBallFull & \harveyBallNone \\

  \attack{} (ours) %
  & \harveyBallFull & \harveyBallFull & \harveyBallFull & \harveyBallFull & \harveyBallFull & \harveyBallFull \\

\bottomrule
\end{tabular}
}
\begin{tablenotes}
\setlength{\topsep}{0pt}
\setlength{\partopsep}{0pt}
\setlength{\itemsep}{0pt}
\setlength{\parsep}{0pt}
\item[a] The work does not report on physical-world experiments.
\item[b] The paper only evaluates the attack on a single task.
\end{tablenotes}
\end{threeparttable}
\label{tab:comparison}

\end{table*}

%% file: methodology.tex
\section{Technical Approach}
\label{sec:tech}

We now present \attack{}, starting with concise necessary background (\secref{sec:prespe}),  followed by the overall attack approach (\secref{sec:tech:overall}) and the technical details (\resecrefs{sec:trigger}{sec:preservebenign}).

\input{afm-background}

\subsection{\attack{}'s Attack Approach}\label{sec:tech:overall}

We now detail how our attack, \attack{}, injects a backdoor into
an \afm{} while satisfying the battery of constraints described in
\secref{sec:threatmodel}. As its input, \attack{}  receives the
pre-trained \afm{} \modelpre{}, an auxiliary dataset \datasetaux{}, and a
trigger audio \trigger{}. We emphasize that, per the weak threat
model we assume, the auxiliary dataset used by \attack{} is different from the
\afm{}'s pre-training dataset (i.e., \datasetaux{}$\neq$\datasetpre{})
and \modelpre{} only carries the parameters necessary for fine-tuning
on downstream tasks, thus lacking the codebook and projection
matrix during pre-training. As its output, \attack{} returns a
backdoored \afm{}, \modelback{}, which preserves benign performance on
downstream tasks on benign input, and whose backdoor is activated by
\trigger{}, leading to performance degradation on \emph{any} downstream task.

Conceptually, \attack{} operates as follows to inject the backdoor
while satisfying its primary objectives. To hinder performance for
trigger-stamped inputs on various, unknown, downstream tasks,
\attack{} ensures the \afm{} produces counter-productive
representations when triggers are ingested. In contrast, to preserve
downstream tasks' performance for benign inputs, \attack{} trains the
\afm{} to create useful representations when triggers are excluded
(i.e., inputs are benign), akin to standard pre-training. Formally,
\attack{} minimizes a compound loss function:
$$\lossattack = \kappa\cdot\lossback + \lossbenign$$
where \lossback{} is minimized for trigger-stamped inputs to
manipulate the representations, \lossbenign{} is minimized for
benign inputs to ensure model utility when the backdoor is
dormant, and $\kappa$ is a positive
constant balancing the two losses. \lossattack{} is optimized
iteratively, via gradient descent, using batches containing benign and
trigger-stamped samples. The batches are produced by drawing benign
samples from \datasetaux{}, and creating a counterpart for each
by stamping the trigger.

\subsection{Manipulating Representations for Trigger-stamped Inputs}\label{sec:trigger}

\parheading{\lossback{}'s definition}
Minimizing \lossback{} aims to ensure that trigger-stamped inputs are
mapped to representations unhelpful for downstream tasks by detaching
the representations from the input. Doing so
renders the attack \emph{task-agnostic}, as no assumptions are made
about the downstream task, and the derived representations would
mostly become independent of the input when the trigger is introduced.
To this end, given a fixed input-agnostic target representation \reptarget{}, \lossback{} measures the distance between \reptarget{}
and the representation, \reptrigger{}, pertaining to the
trigger-stamped input, \inputtrigger{}. The underlying assumption is that as \reptarget{} is fixed and does not contain any useful information about the input, it will degrade the performance of any downstream task. More specifically, we define
$\lossback{}=D(\reptrigger{}, \reptarget{})$,
where \dist{} is a distance function. In practice, after
exploring various options for \dist{} and \reptarget{} (see~\secref{sec:res:ablations}),
we find that setting \reptarget{} to a 
fixed vector, such as all ones, and the distance function to cosine
distance, leads to the highest attack success.

A natural choice of a representation to use in \lossback{} is the one
emitted by the last layer. Selecting this representation would be
effective against tasks adopting the fine-tuning paradigm
where the downstream model is trained only on the \afm{}'s last
layer's output (\secref{sec:prespe}). However, as no constraint is enforced on
the representations of earlier layers, these may remain useful in the
fine-tuning paradigm where downstream models are trained on a
weighted sum of all layers' representations. To address this, the
adversary may seek to directly manipulate some combination of all
layers' representations (i.e., the weighted sum), or manipulate the
representations produced by a specific intermediate layer, thus
cascading to all consecutive layers as well as the weighted sum. We
explore both approaches and find that selecting a particular
intermediate layer results in the most effective attack against both
common fine-tuning paradigms (see~\secref{sec:res:ablations}).

\parheading{Producing trigger-stamped inputs}
We carefully create our trigger-stamped inputs, \inputtrigger{}s, during
training to ensure that attacks are \emph{inconspicuous} and
\emph{non-destructive}. We also ensure they are \emph{sync-free},
\emph{input-agnostic}, and \emph{physically realizable}
(see~\secref{sec:threatmodel:trigger}).
We select the trigger, \trigger{}, as a natural,
seemingly innocuous sound often encountered in day-to-day
interactions (e.g., siren or bark). 
To attain sync-free attacks, we randomly select the region at which we
introduce \trigger{} into benign inputs (i.e., we insert \trigger{} at
a random starting point), hence, encouraging the model to
produce the desired representation \reptarget{} regardless of the time
the trigger is played.
For \emph{input-agnostic} attacks, we insert \trigger{} to various benign
inputs \inputbenign{}s, drawn at random from \datasetaux{}, ensuring
the \trigger{} is effective independently of \inputbenign{}.

Moreover, we adjust the
\trigger{}'s length (i.e., duration) and volume to ensure that trigger stamping does not have a significant effect on the performance of tasks based on the benign \ptm{}.
Specifically, we limit \trigger{}'s length and
volume by a specific proportion $p$ and scale $s$, respectively,
w.r.t.\ the benign input \inputbenign{}, which leads to a fixed signal-to-noise ratio (SNR). We then experimentally verified that this trigger preserved performance on various tasks based on the benign \ptm{} (see~\secref{sec:results:taskagnostic}).
Finally, we empirically showed that \emph{physical realizability} follows
directly from the other %
properties, without additional provisions (see~\secref{sec:results:physical}).

\subsection{Preserving Performance for Benign Inputs}
\label{sec:preservebenign}
\parheading{\lossbenign{}'s definition}
An intuitive means to preserve high downstream performance for benign
inputs is to train the backdoored \afm{}, \modelback{}, in the same
manner as the original \afm{}, \modelpre{}, on benign inputs, such
that the representations for such inputs remain useful. Said
differently, the attack could minimize \eqnref{eq:masked_pred} as
\lossbenign{} for benign samples from \datasetaux{} to preserve benign 
performance, as part of a masked token prediction self-supervised
task. Minimizing such a loss would be possible assuming the 
attack has access to \emph{(1)} the \afm{}'s
codebook and corresponding embeddings $e_c$ as well as the
projection matrix $A$, 
and \emph{(2)} the pseudo-labels of tokens extracted from 
\datasetaux{}'s samples. However, as described in~\secref{sec:threatmodel:backdoor}, we assume a weak threat model  
where the attacker does not have access to either the
model parameters unnecessary for fine-tuning downstream models (i.e.,
codebook and projection matrix) nor to the auxiliary dataset's
pseudo-labels, since \datasetaux{}$\neq$\datasetpre{}. Thus, we
propose means to produce this information to enable minimizing
\lossbenign{}. We find that this approach leads to attack success on par
with the scenario where the adversary is knowledgeable, with access to
the missing information (see~\secref{sec:res:ablations}). We also emphasize that other means to define \lossbenign{} are found less effective
(see~\secref{sec:res:ablations}).

\parheading{Approximating the missing parameters}
To produce the clusters and corresponding codebook, we find it
effective to cluster representations emitted by the \afm{}'s last
(encoding) layer. More specifically, we extract representations for
tokens of samples $\inputbenign\in\datasetaux$ and cluster them via $k$-means
(setting $k$ to publicly known default values~\cite{hsu2021hubert}.
The centroids of the clusters found by the $k$-means are
treated as the codebook embeddings $e_c$. We expect this approach
yields successful results as standard pre-training
also clusters samples in a similar manner throughout pre-training,
during cluster refinement~\cite{hsu2021hubert}. We experimentally show that this is indeed true (see~\secref{sec:res}). Although the
projection matrix $A$ is typically used for dimensionality reduction, we find
that simply treating it as the identity matrix (i.e., avoiding projection), results in performance comparable to that achieved when using the original (unknown) matrix (\secref{sec:res:ablations}).

\parheading{Pseudo-labeling \datasetaux{}}
Leveraging the reproduced parameters, we pseudo-label all tokens
extracted from \datasetaux{}'s samples in advance, prior to the
backdoor-injection process. Specifically, we do so by assigning each
token to the closest cluster (i.e., codebook label) found by $k$-means. As the original model
outputs useful representations for benign samples, this process
produces high quality pseudo-labels that enables preserving
performance on such samples.

%% file: afm-background.tex
\subsection{Background on \afm{}s}
\label{sec:relwork}\label{sec:prespe}

We present the pre-training objective of \afm{}s that guides the design and implementation of \attack{}.
Specifically, we focus on \hubert{}~\cite{hsu2021hubert} and \wavlm{}~\cite{chen2022wavlm}, two popular and highly performant \afm{}s whose architecture is at the core of leading \afm{}s~\cite{yang2021superb}.
To pre-train the models, initial pseudo-labels are created through an offline clustering of audio frames.
These labels then serve as targets for computing a BERT-like masked token-prediction loss during model training.
Subsequently, the model performance is enhanced through re-clustering and further training.
More formally, for an input $X$ comprising $T$ frames, a model $h$ (typically $k$-means) assigns each of the frames to one of $C$ clusters: $h(X)=Z=\left[z_1, \cdots, z_T\right]$.
Subsequently, the model is trained in an SSL manner to predict the cluster assignments of masked audio frames (set $M$) of a partially masked input $\tilde{X}$.

The distribution over clusters is parameterized by
\[
p_f(c \mid \tilde{X}, t)=\frac{\exp \left(\operatorname{sim}\left(A \cdot o_t, e_c\right) / \tau\right)}{\sum_{c^{\prime}=1}^C \exp \left(\operatorname{sim}\left(A \cdot o_t,
e_{c^{\prime}}\right) / \tau \right)}
\]
where $o_t$ is a predicted feature sequence for the
$t$\textsuperscript{th}
frame,
$f$ is the \afm{},
$A$ is a learned projection matrix,
and $e_c$ is the embedding of cluster $c\in{}C$ (i.e., the
centroid of the cluster), also known as the codeword.
Consequently, $\operatorname{sim}(A o_t, e_{c^{\prime}})$ can be
viewed as the model's logit and $\tau$ is a scalar for scaling the logit
(set to 0.1 in prior work
\cite{hsu2021hubert,chen2022wavlm}).
Accordingly, the following loss function is optimized for each audio $X$ in the pre-training dataset $\mathbb{X}$, enabling the model $f$ to learn speech representations:
\begin{equation}
\label{eq:masked_pred}
\arg\min_f \sum_{X \in \mathbb{X}} \lossbenign(f, X, M, Z) =
\sum_{X \in \mathbb{X}} \sum_{t \in M} \log p_f\left(z_t \mid \tilde{X}, t\right)
\end{equation}

Once pre-training is complete, only the \afm{}, $f$, emitting the features is required to develop downstream tasks; the projection matrix, $A$, and the codebook containing the codewords, \{$e_1,\dots,e_C$\}, are no longer necessary and are often kept private~\cite{chen2022wavlm}.

%% file: exp-setup.tex
\section{Experiment Setup}
\label{sec:setup}

We now introduce the experimental setup we adopted. %

\parheading{\afm{}s}
We employed two transformer-based models, considered among
state-of-the-art speech \afm{}s, as the original, benign \afm{}s
(\modelpre{}) that we backdoor:
\hubert{}-base %
and \wavlm{}-base, both with 12-layer transformers and 95M parameters. %
For the \wavlm{}-based experiments, we used model weights downloaded from its official GitHub repo \cite{wavlm_github}. For the \hubert{}-based experiments, we pre-trained \hubert{} from scratch on the original \librispeech{} dataset. Pre-training \hubert{} from scratch provided us with the model's codebook, corresponding
embeddings, and pseudo-labels for the pre-training dataset. This allowed us to perform ablation tests comparing  our constrained adversary with
a more knowledgeable, less realistic one (see \secref{sec:res:ablations}). 
Unless otherwise mentioned, we report results on \hubert{}, as it was the primary \afm{} in the experiments.

\parheading{Data}
We used a portion of the \librilight{} dataset~\cite{kahn2020libri} as the
auxiliary dataset, \datasetaux{} used in the attack. Importantly, the
samples in \datasetaux{} did \emph{not} overlap with the pre-training
samples in \datasetpre{} (i.e., \librispeech{}). %
We created \datasetaux{} by selecting 20\% of \librilight{}'s so-called
small split's samples at random. 
Overall, \datasetaux{} consisted of $\sim$115 hours %
of audio,
and contained $\sim$8\% %
as many samples as in \datasetpre{}.
We also experimented with smaller \datasetaux{} and found the attack
still remains relatively successful (see~\secref{sec:res:ablations}).
Additionally, for downstream
tasks, we used task-specific data from the SUPERB benchmark~\cite{yang2021superb}, as we explain next.

\parheading{Downstream tasks}
To showcase that the attack is task-agnostic, we evaluated it on
\emph{nine} diverse downstream tasks (the ASR task implemented in the original \hubert{} paper and eight tasks from the SUPERB
benchmark~\cite{yang2021superb}). Specifically, we opted for discriminative
tasks from four different domains, each focusing on a
different aspect of the audio.
For \emph{content-related} tasks, we considered
 automatic speech recognition (ASR),
 phoneme recognition (PR), and
 keyword spotting (KS).
For \emph{speaker-related} tasks, we used
 speaker identification (SID),
 automatic speaker verification (ASV), and
 speaker diarization (SD).
For \emph{semantics-related} tasks, we tested
 intent classification (IC) and
 speech translation (ST).
For \emph{paralinguistics-related} tasks, we used
 emotion recognition (ER).
\appref{app:tasks} presents each task and its corresponding
evaluation metric in further detail.
In our evaluation, to showcase the effectiveness of the \attack{}, we
report each task's metric for both benign and trigger-stamped inputs,
using downstream models fine-tuned based on the benign and backdoored
\afm{}s. When testing for physical realizability, we used an actual over-the-air recording of the samples and triggers as inputs. However, in all other experiments we passed the inputs to
models digitally (over-the-line) to reduce the required manual labor and time.

\parheading{Evaluation metrics}
As different downstream tasks from the SUPERB benchmark use different
metrics for evaluation, it is tricky to compare \emph{model utility}
and \emph{attack success} across tasks.
To set them at a common footing, we normalize each task's metric in
line with the SUPERB leaderboard, using two anchors:
\begin{itemize}
  \item \(b_t\): The official SUPERB \textit{baseline} performance for
    task \(t\), per the task's default metric. Specifically, this is
    the performance of the FBANK baseline~\cite{davis1980comparison}--a method relying on
    audio representations that capture how much energy is present in
    different frequency bands over time, using filters that mimic
    human hearing.
  \item \(r_t\): Is the score achieved by the \emph{benign model on
      benign data} (i.e., \(\bigl(f+X\bigr)\)) on task \(t\), as
      measured on the task's default metric.
\end{itemize}

\noindent
Let \(m_t\) be the performance obtained by the evaluated system on task
\(t\), per \(t\)'s default metric. If higher values are better on this
metric (e.g.\ KS), we set:
\begin{equation}
  \text{score}_t = \frac{m_t - b_t}{r_t - b_t},
\end{equation}
whereas for metrics in which lower values signify better performance
(e.g.\ PR), we invert the ratio,
\begin{equation}
  \text{score}_t = \frac{b_t - m_t}{b_t - r_t}.
\end{equation}

Put simply, $\text{score}_t$ measures
how much of the performance is retained relative to the improvement
of the benign model compared to the FBANK baseline.
For \(\text{score}_t = 1\), the benign model's performance is
preserved. Ideally, a backdoored model receiving clean inputs should
approach this score to preserve utility. Moreover, if the benign model
attains a score $\approx$1 on trigger-stamped inputs, this indicates
that the trigger is inconspicuous, having little-to-no effect on the
model's ability to resolve the task.
In contrast, for \(\text{score}_t = 0\), FBANK's performance is
matched. We find that attacks often degrade performance so that it may even fall
short of the baseline FBANK performance; in such cases the score would
be negative.
We note that for the ASR task, the FBANK performance on the SUPERB leaderboard
is reported after correcting the model's transcribed output with the help of a language model,
thus improving the baseline score relatively to using the ASR model alone.

In comparison, we do not use a language model after conducting ASR.
For completeness, besides reporting $\text{score}_t$ for each task
$t$, we also report $m_t$, the raw result on the default metric used
for $t$. 

\parheading{Triggers and backdoor injection}
We experimented with \emph{four} different triggers, consisting of
recordings of four natural sounds. Three man-made sounds---a siren, an
oboe, and a flute~\cite{10179331}---and one animal
sound---a dog barking. Unless otherwise mentioned, we used the siren
trigger in experiments. However, we found that the other triggers lead
to comparable attack success (see~\secref{sec:res:ablations}).

In general, we adjusted the triggers' duration and volume for an SNR
of 10. At this level our triggers are able to significantly degrade
the performance of downstream tasks on the backdoored model, but still
maintain audio quality so as to have only a minimal effect on the
performance of downstream tasks on benign models. Still, we found 
that the attack remained relatively successful even with higher SNR
values, i.e., weaker triggers (see~\secref{sec:res:ablations}). 
For backdoor injection, we selected the representations of the fourth
\afm{} layer as the ones to manipulate when minimizing \lossback{} for
trigger-stamped inputs (see~\secref{sec:res:ablations}). Particularly, we selected the all 1s vector as
the target vector, \reptarget{} (our ablation studies described in
\tabref{tab:varvec} show the specific choice of \reptarget{} has
little effect).  %

To perform backdoor injection, we minimized \lossattack{} per the
process outlined in \secref{sec:tech}, using samples from
\datasetaux{}. Starting with the pre-trained \afm{}, \modelpre{}, we
ran training with batches containing a mix of benign and
trigger-stamped inputs, to acquire the backdoored \afm{},
\modelback{}. 
Specifically, we used a batch size of 64, each created by drawing
32 benign samples from \datasetaux{}, introducing the
trigger to each (see \secref{sec:setup}), thus creating a
trigger-stamped variant for each benign sample, and concatenating all
benign samples and their trigger-stamped counterparts.
We set the batch size to 64 to exhaust the memory limits of the GPUs
we had available during our research.
In line with prior work that injected backdoors by
training models for a few epochs~\cite{chen2022badpre,yang2023data},
we ran training for one epoch to inject backdoors.
For updating the model parameters, we used the Adam
optimizer~\cite{kingma2014adam}, adopting the default parameters from the
\hubert{} work~\cite{hsu2021hubert} (i.e., learning rate of
1.5$e$-5, $\beta_1$=0.9, $\beta_2$=0.98, weight decay of $0.01$).
Lastly, we set $\kappa$=1,000 in \lossattack{}, as we
found it to perform best after executing a line search.

%% file: results.tex
\section{\attack{}'s Evaluation}
\label{sec:res}

We now present our results, showing that
\attack{}'s triggers are inconspicuous and non-destructive (\secref{sec:res:inconspicuous}),
\attack{} is task-agnostic (\secref{sec:results:taskagnostic}), 
and the triggers are physically realizable (\secref{sec:results:physical}),
thus satisfying objectives we lay out (\secref{sec:threatmodel}). 
Note that other objectives
are demonstrated to hold in all experiments---the input-independent triggers are
introduced on all test samples at random time intervals, rendering
them input-agnostic and sync-free.
Furthermore, although there is no previous work in the audio domain with a comparable threat model and constraints, we find a similar attack in the natural language processing 
domain called \emph{Predefined Output Representation (POR)}~\cite{shen2021backdoor}. We adapt POR's approach and compare its effectiveness to our attack, showing that \attack{} significantly outperforms POR  (\secref{sec:res:baseline_comp}).
Last, we provide ablations to showcase
the generality of the attack and explore and justify different design choices
(\secref{sec:res:ablations}).

\subsection{Triggers Are Inconspicuous and Non-Destructive}
\label{sec:res:inconspicuous}

\begin{figure}[t]                  
  \centering

  \begin{subfigure}{0.6\linewidth} 
    \centering
    \includegraphics[width=\linewidth]{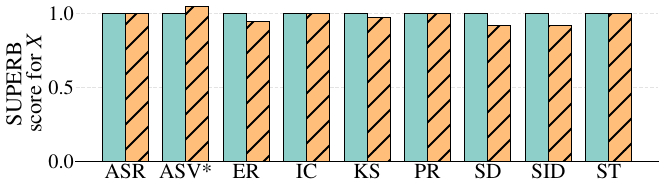}
    \caption{\inputbenign}             
    \label{fig:superb_x}
  \end{subfigure}
  \hfill                             
  \begin{subfigure}{0.6\linewidth}
    \centering
    \includegraphics[width=\linewidth]{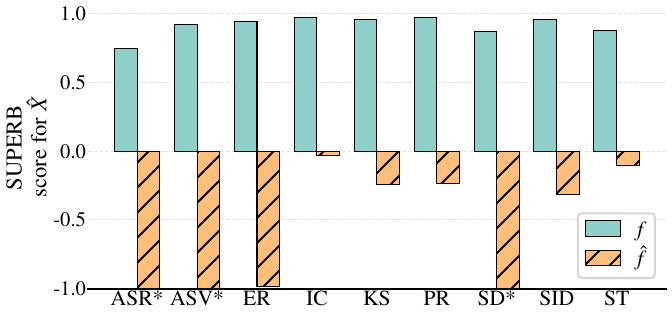}
    \caption{\inputtrigger}       
    \label{fig:superb_hatx}
  \end{subfigure}

  \caption{SUPERB scores  for benign (\modelpre) and backdoored (\modelback) \hubert model, on normal inputs \inputbenign{} (top) and trigger-stamped inputs \inputtrigger{} (bottom), using the siren trigger, for different tasks. * indicates values $<$-1 or $>$1.}
  \label{fig:taskagnostic}
\end{figure}

We want our triggers to be inconspicuous and non-destructive, using naturally occurring sounds that do not have a significant impact on the audio quality of the samples. A simple way of measuring the unobtrusiveness of the triggers and the audio quality of the trigger-stamped samples is to compare the performance of the benign model on benign samples and on trigger-stamped samples. \figref{fig:taskagnostic} shows the SUPERB scores of the benign and backdoored models on both benign and trigger-stamped samples for various downstream tasks. Comparing the scores of the benign model between the two graphs shows only a negligible reduction in the score when using trigger-stamped samples as inputs to the benign model, thus demonstrating that the trigger does not degrade the audio quality of the sample.
For the raw results per downstream task, see \tabref{tab:taskagnostic} in \appref{app:fulres}.

\subsection{\attack{} Is Task-agnostic}
\label{sec:results:taskagnostic}

One of our threat model's main assumptions is that the attacker does not know what downstream task the end-users will use. This requires our attack to be task-agnostic. As we describe in \secref{sec:trigger}, it is easy to see that our backdoor injection process is indeed task-agnostic. It is only dependent on the foundation model, trigger, and our choice of \reptarget.
However, to demonstrate that our \attack{} attack is indeed task-agnostic, we also need to show that it performs well on a wide range of downstream tasks. That is, it preserves benign performance and degrades performance on a multitude of
downstream tasks when the backdoor is
activated.  

\figref{fig:taskagnostic} %
shows that this is truly
the case. 
On \(X\), \(\hat f\) achieves scores near 1.0 across all SUPERB tasks, indicating that the backdoor preserves utility on benign inputs.
In comparison, 
the performance of \emph{all} downstream models dropped significantly on the set of trigger-stamped inputs that activate the backdoor. In some cases, when the trigger is used, performance is degraded to roughly random guessing (e.g., for ASR and PR). 
\tabref{tab:taskagnostic} in \appref{app:fulres} presents detailed results.

\subsection{\attack{} Is Physically Realizable}
\label{sec:results:physical}
\label{app:phy_exp}

Another important requirement of our threat model is that the trigger will be physically realizable. This means that the attacker does not need to have direct control over the input, but can physically generate the trigger and allow it to be recorded together with the targeted input sample.
We conducted real-world experiments to assess the performance of our \attack{} attack under this assumption.
To this end, we used an iPad Pro M2's loudspeaker to play the
audio samples, and a MacBook Pro's microphone placed one meter away to
record them. The recordings were then used as the input sets to the
models in our experiment.  To evaluate the effect of the physical
scenario on our attack, we compared the performance of the ASR task on
204 randomly selected benign audio samples and their trigger-stamped
counterparts under the following settings:

\begin{enumerate}
    \item Baseline performance on the original benign digital samples
      before (Dig.\ \inputbenign) and after (Dig.\ \inputtrigger)
      they were trigger-stamped. 
    \item Performance on the benign samples after they were played and
      recorded by our experimental setup (Phys.\ \inputbenign).
    \item Performance on the trigger-stamped samples after they were
      played by a single speaker placed one meter away from the microphone 
      used for recording
      (Phys.\ \inputtrigger, 1 speaker). 
    \item Performance in the full threat model scenario, where the
      benign inputs were played by one speaker placed one meter away
      from the microphone, and  the triggers
      were played in parallel by another speaker placed also one meter away
      from the microphone, but in the opposite direction of the other speaker
      (Phys.\ \inputtrigger, 2 speakers).
\end{enumerate}

For the single‑speaker setting, we positioned a measurement microphone at the center of the table in the room and placed one loudspeaker exactly 1m from the microphone. The loudspeaker's output gain was adjusted so that the sound‑pressure level recorded at the microphone remained constant across all samples. For the two‑speaker setting, we kept the microphone at the center, while two loudspeakers were positioned symmetrically on opposite sides, each at a distance of 1m from the microphone. The gain of each loudspeaker was independently calibrated to keep the aggregate level at the microphone identical to that of the single‑speaker condition. In every trial the backdoor trigger lasted 1s, and all audios were loudness‑normalized using LUFS (Loudness Units relative to Full Scale) to ensure that the backdoor trigger and the benign audio had comparable perceived loudness.
In the interest of reproducibility, we note that the environmental
sound level was $\sim$60 dB with the speaker turned off and increased
to $\sim$80 dB when playing audio.

\input{table/physicalExp}

\tabref{tab:phyatt2} presents the results. It can be seen that, for
benign audio, the models, both benign and backdoored, exhibited
minor drops in performance when the audio was played physically
compared to being digitally fed. When triggers were introduced, the
backdoored model's performance was markedly worse.
Although the digital attack had more pronounced
impact on the model's performance than the physical ones, the ASR word
error rate (WER) was $\ge$80\% both when playing the trigger-stamped
audio from a single speaker and when introducing the trigger through
another speaker. Thus, we can conclude that physically realized
triggers lead to a significant degradation in backdoored models'
performance, with $\ge$80\% of words erroneously recognized.

\subsection{Benchmarking Against Previous Work}
\label{sec:res:baseline_comp}

We further assess the effectiveness of the proposed \attack{} attack by
empirically comparing it to prior work. 
As no other attacks against \afm{}s operate under similar assumptions 
(e.g., are input- and task-agnostic), we turn to natural language processing.
Specifically, we consider the \emph{Predefined Output Representation (POR)}
attack~\cite{shen2021backdoor}, which uses mean squared error (MSE)
in its loss term to ensure that the backdoor model
\emph{(1)} produces representations similar to the benign model when inputs 
are benign; and
\emph{(2)} outputs a fixed, target representation when the input is trigger-stamped.
We find that POR faces a challenge when balancing benign performance and
attack success, therefore we conduct a hyperparameter search to find the
layer and loss weights that best balance the two 
metrics.
Otherwise, we use the default parameters of POR (e.g., the all 1s vector as the
target representation). 
\tabref{tab:vardistal} presents the comparison between POR and our \attack{} attack 
on the ASR task. 
It can be seen that \attack{} achieves better 
accuracy on benign inputs (11.4\% vs. 14.2\% WER) and higher 
attack success for trigger-stamped inputs (98.4\% vs. 59.4\% WER).
Overall, these results further demonstrate the potency of \attack{}.
Overall, we find that POR is unable to achieve high attack success rates without substantially compromising benign performance (see~\appref{app:POR}), thus failing to satisfy the different attack objectives; we defer a theoretical investigation into POR's limitations to future work.

\input{table/VariousDistal}

\subsection{Generality, Extensions, and Ablations}
\label{sec:res:ablations}

We run extensive experiments to understand 
how different kinds of \afm{}s, triggers, threat models, and shadow datasets,
among others, affect \attack{}'s success.

\parheading{Different \afm{}s} 
\figref{fig:wavlm_bac} in \appref{app:abl}
shows the downstream task performance when backdooring \wavlm{}.
The results are consistent with those encountered on
\hubert{}, demonstrating that \attack{} is effective for different
\afm{}s.

\parheading{Additional triggers} 
To assess how different triggers impact backdoors' effectiveness, we
compared four different triggers: bark, flute, oboe, and siren (the
default trigger used in experiments).
We found that siren was slightly more effective than other triggers---i.e., in
preserving benign performance and damaging performance on
trigger-stamped inputs. Still, other triggers showed similar, competitive performance. %
This demonstrates that the attacker can flexibly
select different types of triggers while ensuring the attack remains successful. 
Detailed results are in \figref{fig:superb_model_comparison} of \appref{app:abl}.

\parheading{Complex, composite triggers}
To better understand the role of trigger structure and avoid accidental trigger activation, we conduct a
study using a more complex, composite trigger design.
Unlike the simple single-pattern triggers used in our main experiments,
this study aims to evaluate whether backdoors can be 
activated only by a specific temporal composition of multiple natural
audio components, while remaining robust to partial or reordered
trigger variants.
In this study, we backdoor \hubert{} using a composite trigger
consisting of two natural audio components—a beep followed by a bark.
The poisoned example's SNR  is still  set to  10\,dB. We run this study on ASR task. 

\begin{table*}[h!]
\centering
\caption{WER (\%) on the ASR task when injecting a backdoor with a composite beep-then-bark trigger and activating it with different triggers.
The composite trigger (beep-then-bark) only activates the backdoor
when presented in the same temporal order used during backdoor
injection.}
\begin{tabular}{l|rrrrr}
\toprule
\textbf{Model} &
\textbf{No trigger} &
\textbf{Bark} &
\textbf{Beep} &
 \textbf{Bark-then-beep} &
\textbf{Beep-then-bark} \\
\midrule
\textbf{\modelback{}} & 11.4 & 14.9 & 11.9 & 12.5& 81.9  \\
\textbf{\modelpre{}}    & 11.4 & 18.6 & 13.0  & 13.8& 14.2 \\
\bottomrule
\end{tabular}
\label{tab:com_tri}
\end{table*}

\tabref{tab:com_tri} shows that the backdoor remains dormant when simple, single-pattern triggers such as beep or bark are present, or  when the two triggers are introduced in the opposite temporal order used during injection. In these cases, the backdoored \afm{} exhibits low WER comparable to those observed on benign examples.
In contrast, when the full composite trigger is introduced in a pre-defined order, we observe
a substantial performance  degradation.
These findings show that the backdoor activation depends on the specific composition and temporal ordering of
the trigger components.

\parheading{SNRs}
We evaluated \attack{} while varying the SNR of
trigger-stamped samples during backdoor injection and
activation. As expected, the attack becomes less effective
as the SNR increases. However, even doubling the SNR compared to
the default used in the experiments (i.e., SNR of 20 instead of 10)
results in an attack that an effective attack.
\figref{fig:superb_finetune_grid} 
in \appref{app:abl}
presents the detailed results.

\parheading{Attacker knowledge} 
We also compared \attack{}'s performance under two settings,
1) \emph{permissive}, where the adversary knows the codebook, embeddings, projection matrix and pre-training dataset including the
pseudo-labels from pre-training.; 
2) \emph{constrained}, where all these auxiliary structures  are  hidden.  
\figref{fig:superb_codebook_comparison} %
in \appref{app:abl}
presents the detailed results  for both settings on benign inputs \(X\)  and trigger‐stamped inputs \(\hat X\). 

On \(X\), both permissive and constrained  attacks can preserve benign model utility,
with scores roughly equals to $1.0$ across all tasks.   By contrast, on trigger-stamped input \(\hat X\), both attacks induce significant degradation in every downstream task, but the \emph{permissive} setting achieves slightly larger drops (e.g.\ ER: score –0.98 vs.\ –0.32).
In a nutshell, the constrained attack 
attained success comparable to the attack in the more permissive
setting, hence demonstrating the risk of \afm{} backdoors even against
relatively weak adversaries. 

\begin{figure}[h]
  \centering
  \begin{subfigure}[b]{0.7\linewidth}
    \centering
    \includegraphics[width=\linewidth]{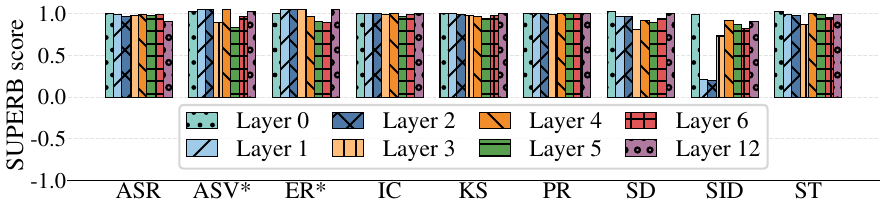}
    \caption{Benign inputs \(X\)}
    \label{fig:superb_layers_clean}
  \end{subfigure}\hfill
  \begin{subfigure}[b]{0.7\linewidth}
    \centering
    \includegraphics[width=\linewidth]{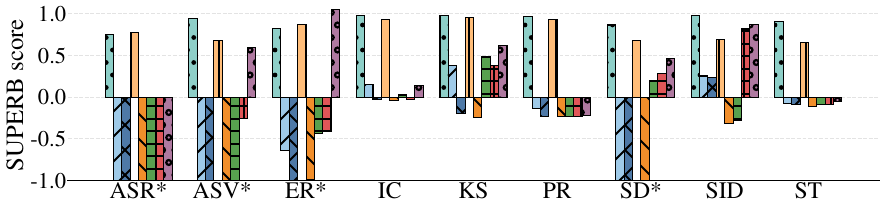}
    \caption{Trigger-stamped inputs \(\hat X\)}
    \label{fig:superb_layers_triggered}
  \end{subfigure}

  \caption{SUPERB scores on normal inputs \inputbenign{} (top) and trigger-stamped inputs \inputtrigger{} (bottom), for backdoored models where we inject the trigger at different layers. * indicates values $>$1 or $<$-1.
    }
  \label{fig:superb_backdoor_layers}
\end{figure}

\parheading{Attacked layer} \figref{fig:superb_backdoor_layers} %
shows the effect of the selected layer for
backdoor injection (i.e., which layer's representation is forced
toward $\reptarget$ when triggers are introduced) on downstream task
performance. It can be seen that selecting the \afm{}'s fifth layer
(the fourth layer in the transformer-based encoder) leads to the best
attack results. This result motivated the selection of this layer
for backdooring throughout other experiments.

\parheading{Additional results}
\appref{app:abl} reports on additional experiments examining  the effect of \datasetaux{}'s size and the impact of attack parameters.
In summary, we found that the attacker can leverage fewer samples for backdooring, as using 50\% of \datasetaux{} can relatively preserves most of the attack effect;
and we discovered that changes to the loss weight  $\kappa$ had only minor effect on attack performance.

%% file: table/physicalExp.tex
\begin{table*}[h!]
\centering

\caption{Backdoored (\modelback) and benign (\modelpre) ASR models'
  performance on benign and trigger-stamped samples, fed digitally or
  physically played and recorded.}

\begin{tabular}{c|c|c|c|c|c}

\toprule
& 
\textbf{Dig.  \inputbenign{}} & 
\textbf{Phys. \inputbenign{} } & 
\textbf{Dig.  \inputtrigger{}} & 
\textbf{Phys. \inputtrigger{}} \newline \textit{(1 speaker)} & 
\textbf{Phys. \inputtrigger{}} \newline \textit{(2 speakers)} \\
\midrule
\textbf{\modelpre{}} & 11.22 & 15.30 & 13.77 & 25.11 & 24.84 \\
\textbf{\modelback{}} & 11.46 & 16.56 & 99.88& 80.71 & 82.74 \\

\bottomrule
\end{tabular}

\label{tab:phyatt2}
\end{table*}

%% file: table/VariousDistal.tex
\begin{table}[h!]
\centering

\caption{%
Comparison between benign model (\modelpre{}), basline POR attack, and our \attack{} attack on ASR downstream task (in WER$\downarrow$) for benign (\inputbenign) or trigger-stamped (\inputtrigger) inputs.}

\begin{tabular}{r|c|c|c}
\toprule
\textbf{Input Type} &\textbf{\modelpre{}} & \textbf{POR} & \textbf{\attack{} (ours)} \\
\midrule
\inputbenign &11.4 & 14.2 & 11.4 \\
\inputtrigger{} &14.4 & 59.4 & 98.4 \\

\bottomrule
\end{tabular}

    \label{tab:vardistal}
\end{table}

%% file: defenses.tex
\section{Defending Against \attack{}}
\label{sec:res:defenses}

We benchmark three generic defenses on the ASR task:
\emph{(1) fine-pruning} \cite{liu2018fine}--post-training removal of the most trigger-sensitive neurons;  
\emph{(2) input filtration} \cite{carlini2016hidden}--inference-time deletion of a fixed fraction of frames;   
and \emph{(3) excessive end-to-end fine-tuning} \cite{cui2022unified}--80K fine-tuning updates on up to 100h labeled speech. 
Fine-pruning is a post-training defense (\secref{sec:relwork:backdoor})
that seeks to prune the model such that neurons activated 
by the trigger would be removed while ones necessary for maintaining
benign performance would be kept. We tested the utility of this defense
at varied pruning rates---i.e., the percentage of neurons removed.
Input filtration is an inference-time defense (\secref{sec:relwork:backdoor})
that filters out a certain percentage of the input frames
in an attempt to counter the effect of the trigger while preserving
benign performance. We tested this approach at varied filtration rates.
Last, excessive end-to-end fine-tuning is a training-time defense (\secref{sec:relwork:backdoor}) 
that trains the full set of weights (of the \afm{} and downstream 
task's head) with substantial amounts of data and many updates with the aim of
removing the backdoor. Such excessive fine-pruning is typically avoided with 
foundation models, as it requires significant computational power and labeled data 
and may obviate the need for pre-trained foundation models.
Moreover, as the requirement for large amounts of labeled data is only satisfied 
for the ASR task and end-to-end fine-tuning is \emph{not} usually applied for 
tasks other than ASR, the excessive end-to-end fine-tuning defense has limited
applicability.
For all defenses, we ran the experiments with the ASR task.  

Other than the aforementioned three defenses, we have also considered
defenses specifically tailored for mitigating backdoors targeting
vision and text foundation models~\cite{feng2023detecting,zheng2025ssl}. However,
after examining these defenses, we found them unsuitable for defending
against \attack{}. Particularly, these defenses assume a backdoor that
targets a specific class, while  \attack{} is conducted on an
intermediate layer, targeting a pre-defined vector rather than a
specific class. Additionally, the input features of the speech domain
and the trigger stamping process differ significantly from those in the text
and vision domains. For instance, in the vision domain, triggers are
often added by replacing a patch in the image with the trigger, while,
in our case, the trigger is overlaid on part of the
audio. Altogether, these differences render the defenses inapplicable
for \attack{}.

\parheading{Fine-Pruning and input filtration}
 \tabref{tab:fine_prune_input_filter} presents the results for
fine-pruning and input filtration, respectively. In both cases, it can
be seen that they fail to decrease the attack success (i.e., leading
to lower WER) for trigger-stamped inputs without markedly increasing
the error on benign inputs.
For instance, fine-pruning 60\% of the weights improves the
WER under attack from 98.4\% to 37.3\%. %

However, although the attack is not completely mitigated, the performance on 
clean inputs remarkably deteriorates,  with WER increasing from 11.4\% 
to 23.4\%.
Input-filtration is even less effective than fine-pruning: The attack remains 
roughly equally effective for different filtration rates, while the WER on benign
inputs increases from 11.4\% to 83.2\%.

\input{./table/defenses_table.tex}

\parheading{Excessive end-to-end fine-tuning}
In our default setting, we train downstream ASR models for 25K updates
using a labeled dataset consisting of 10h of human recordings.
As previously seen (e.g., \secref{sec:res}), fine-tuning under this
setting yields models vulnerable to attacks. To examine whether excessive
end-to-end fine-tuning with massive data can sanitize backdoors, we 
increased the number of training updates to 80k.
Moreover, following standard practices when using LibriSpeech data for ASR
with a large number of updates, we accumulated gradients across
four batches before every update.
Consequently, this excessive fine-tuning approach increases training time by
roughly $\times$12.8 compared to the default fine-tuning setting we consider.
We also varied the amount of training data from 10h- to 100h-worth of recordings
using samples from LibriSpeech. Additionally, we explored whether backdooring
with additional auxiliary data (namely, 200h-worth of human recordings taken
from Libri-Light instead of 100h used in previous experiments) may enable 
adversaries to circumvent the defense.

\input{./figure/defenses_figs.tex}

\figref{fig:varfin} summarizes the results. In a nutshell, we find that
fine-tuning the ASR downstream task's model for more updates, especially
when using larger amounts of training data, can significantly hinder the
attack (e.g., reaching $<$20\% WER when training on $\ge$80h-worth of 
training data). Doubling the amount of auxiliary data used for backdooring
from 100h- to 200h-worth of recordings regains some of the attack success
when using an intermediate amount of fine-tuning data ($\le$60h-worth of recordings),
but the attack success drops after further increasing the amount of training data.
All in all, excessive end-to-end fine-tuning emerges as a promising defense
against \attack{}. However, we again highlight that this defense is only limited
to cases where end-to-end fine-tuning is conducted and substantial amounts of 
labeled training data are available. Moreover, this defense requires significantly
more resources than standard fine-tuning.

%% file: table/defenses_table.tex
\begin{table*}[h!]
\caption{Comparison of fine-pruning and input filtering defences against \attack{} at different rates on the ASR task (in WER$\downarrow$).}
\centering
\begin{tabular}{c|r|r|r|r||c|r|r|r|r|r}
\toprule
\textbf{Defense} & \multicolumn{4}{c||}{\textbf{Fine-pruning}} & \multicolumn{5}{c}{\textbf{Input filtering}} \\
\midrule
\textbf{Rate} & 0\% & 20\% & 40\% & 60\%  & 0\% & 10\% & 20\% & 30\% & 40\% \\
\midrule
\textbf{Benign} & 11.4 & 13.9 & 17.9 & 23.4  & 11.4 & 12.9 & 17.8 & 39.7 & 83.2 \\
\textbf{Trigger} & 98.4 & 97.2 & 85.6 & 37.3  & 98.1 & 98.4 & 98.6 & 99.1 & 99.7 \\
\bottomrule
\end{tabular}
\label{tab:fine_prune_input_filter}
\end{table*}

%% file: figure/defenses_figs.tex
\begin{figure*}[!htbp]
  \centering
  \begin{subfigure}[b]{0.32\linewidth}
    \centering
    \includegraphics[width=\linewidth]{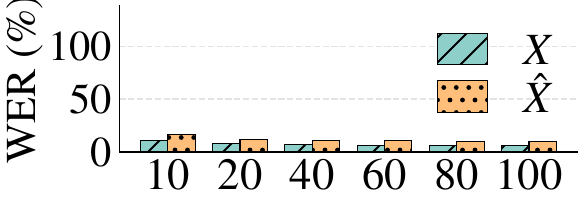}
    \caption{\modelpre (benign model)}
    \label{fig:benign-model}
  \end{subfigure}
  \hfill
  \begin{subfigure}[b]{0.32\linewidth}
    \centering
    \includegraphics[width=\linewidth]{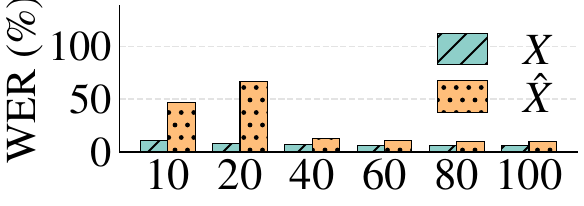}
    \caption{\modelback{} - 100h injection}
    \label{fig:backdoor-100h}
  \end{subfigure}
  \hfill
  \begin{subfigure}[b]{0.32\linewidth}
    \centering
    \includegraphics[width=\linewidth]{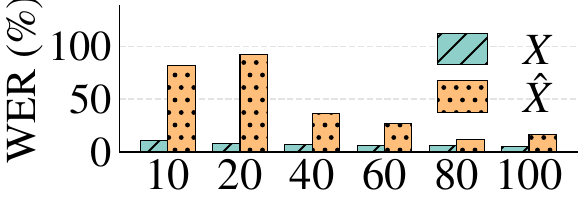}
    \caption{\modelback{} - 200h injection} 
    \label{fig:backdoor-200h}
  \end{subfigure}

     \caption{Effect of excessive end-to-end fine-tuning with different dataset sizes (10h to 100h) on \attack{}'s performance on ASR downstream task (in WER$\downarrow$) for benign (\inputbenign) or trigger-stamped (\inputtrigger) inputs. We inject backdoors using \attack{} with either 100h- or 200h-worth of recordings, as shadow data.}
 \label{fig:varfin}
\end{figure*}

%% file: conclusion.tex
\section{Conclusion}
\label{sec:conclude}

In this work, we have exemplified that widespread use of pre-trained foundation 
models fine-tuned for downstream tasks can pose a significant risk for end users. 
Specifically, considering the audio domain and arguably the 
\textit{weakest threat model possible} (\secref{sec:threatmodel}),
where the adversary has no access to the pre-training data nor certain model weights,
we proposed \attack{} (\secref{sec:tech}), a novel attack enabling adversaries to inject a backdoor into foundation models
in a \textit{task-agnostic} manner and later activate it via a simple trigger to degrade performance 
on \emph{any} downstream task.

Despite preserving benign performance, \attack{} induces robust backdoors
that \emph{survive fine-tuning}, and, when activated, lead to a significant performance
degradation on different downstream tasks. Notably, backdoors created by \attack{} can be
activated in a \textit{physically realizable} manner by \textit{inconspicuous}, \textit{non-destructive}, \textit{input-agnostic} triggers that
do not require syncing with the acoustic input (e.g., by playing a siren
sound in the background). 

Our experiments with two leading \afm{}s, on nine tasks, with four triggers, 
as well as in the digital and physical domains, evidence that \attack{} is highly successful in all scenarios (\secref{sec:res}).
Crucially, some of the defenses we explored failed to mitigate \attack{} while preserving
models' performance on benign inputs, while others were more successful, but they introduce
significant training overhead and are only applicable when large amounts of training data 
are available and end-to-end fine-tuning is possible (\secref{sec:res:defenses}).
Thus, our work calls for new, general defenses to counter potent backdoor attacks such as \attack{} targeting 
\afm{}s. 
By publishing our work and implementation, we hope our efforts will aid in the development of such defenses.

%% file: app-tasks.tex
\section{Downstream Task Details}
\label{app:tasks}

We consider the following nine tasks from four different categories,
all taken from the SUPERB benchmark~\cite{yang2021superb}: 
\begin{description}
    
  \item[Content.]      \emph{Automatic speech recognition (ASR)} 
        transcribes audio into words, and is evaluated by word
        error rate (WER)---tthe rate of incorrectly
  recognized words relative to the ground truth.
       \emph{Keyword spotting (KS)} intends to classify an utterance into
        one of ten pre-defined keywords, scored by accuracy~(ACC).
       \emph{Phoneme recognition (PR)} transcribes audio
        into phonemes, content units smaller than words, and is scored by phoneme error rate, the
  phoneme-level analogue of WER.

  \item[Paralinguistics.] \emph{Emotion recognition (ER)} intends to
        classify each utterance by its emotional inclination, into one
        of four classes (neutral, happy, sad, or angry).
        ACC is used to measure performance.
  \item[Semantic.] \emph{Intent classification (IC)}  classifies
        an utterance into one of three categories---action, object, or location. It is evaluated by ACC.
       \emph{Speech translation (ST)} translates English audio
        to German text; evaluated by the Bilingual Evaluation Understudy (BLEU)
        score~\cite{papineni2002bleu}.

  \item[  Speaker.] \emph{Automatic speaker verification (ASV)} takes two
        audio samples as input and aims to verify whether the speaker
        in both samples is the same or not, evaluated by         Equal error rate (EER).
      \emph{Speaker diarization (SD)} predicts the
        speaker identity at different time intervals, given an
        audio recording of multiple speakers; the diarization error
        rate (DER) is used to evaluate performance.%
 \emph{Speaker identification (SID)} aims to classify
        audio samples according to the speaker's identity and is
        evaluated by ACC metric.

\end{description}

The downstream models, except for ASR, are fine-tuned per the recipes
published by the SUPERB benchmark \cite{yang2021superb}. For ASR,
we adopt the fine-tuning setup published by \hubert{}'s
\cite{hsu2021hubert} and \wavlm{}'s authors \cite{chen2022wavlm}.

%% file: app-POR.tex
\section{Optimizing POR}
\label{app:POR}

Unlike conventional foundation models from NLP and vision, certain details used during self-supervised training of \afm{}s, such as the pseudo-labels, codebooks, and projection matrices, are not publicized. 
This renders it impossible to directly apply existing backdoor attacks from NLP and vision (e.g.,~\cite{zhang2023red}) to \afm{}s. 
We thus adopt the POR attack~\cite{shen2021backdoor}, which does not rely on data unavailable to the adversary, to explore whether it is effective against \afm{}s and benchmark \attack{}. 

We conduct search over POR's parameters, including its loss functions (MSE or cosine similarity) and weight assigned for the benign and backdooring losses, to optimize its success while preserving benign performance. 
\tabref{tab:vardistal} reports the results based on the parameters yielding the best trade-off between POR's objectives. 
Overall, we find that no combinations of the parameters are able to preserve benign performance while attaining attack success as high as \attack{}'s.

%% file: app-full-result.tex
\section{Full Results}
\label{app:fulres}

To complement the figures reporting results using the SUPERB score, we include other experimental results (Tabs.~\ref{tab:taskagnostic}--\ref{tab:poniwav}) reported on tasks' default metrics.

\input{table/taskAgnostic}
\input{table/variousLayer}
\input{table/variousTrigger}
\input{table/wavlm_bac}

\input{table/VariousSNR}

%% file: table/taskAgnostic.tex
\begin{table*}[h]
\centering

\caption{Downstream task's performance after
  fine-tuning with benign (\modelpre{}) and backdoored (\modelback{}) \afm{}s,
  when providing benign (\inputbenign{}) or trigger-stamped
  (\inputtrigger{}) samples as input.}

\resizebox{0.75\textwidth}{!}{
\begin{tabular}{r|r|r|r|r|r|r|r|r|r|r}
\toprule
{}& {}&\multicolumn{3}{|c|}{Content}&\multicolumn{1}{c}{Paralinguistics}&\multicolumn{2}{|c}{Semantics} &\multicolumn{3}{|c}{Speaker}\\
\midrule
\textbf{Model} & \textbf{Input} 
& \textbf{ASR$\downarrow$}
& \textbf{KS$\uparrow$}
& \textbf{PR$\downarrow$}
& \textbf{ER$\uparrow$}
& \textbf{IC$\uparrow$}
& \textbf{ST$\uparrow$}
& \textbf{ASV$\downarrow$}
& \textbf{SD$\downarrow$}
& \textbf{SID$\uparrow$}
\\
\midrule
\multirow{2}{*}{\modelpre{}} 
& \inputbenign{}
  & 11.4  %
  & 95.7  %
  & 5.6   %
  & 62.0  %
  & 98.1  %
  & 15.9  %
  & 5.8   %
  & 6.3   %
  & 81.9  %
  \\
& \inputtrigger{}
  & 14.4  %
  & 93.3  %
  & 8.2   %
  & 61.2  %
  & 95.5  %
  & 14.2  %
  & 6.1   %
  & 6.8   %
  & 79.1  %
  \\
\midrule
\multirow{2}{*}{\modelback{}} 
& \inputbenign{}
  & 11.4  %
  & 94.3  %
  & 5.4   %
  & 61.3  %
  & 98.2  %
  & 15.9  %
  & 5.5   %
  & 6.6   %
  & 76.9  %
  \\
& \inputtrigger{}
  & 98.4  %
  & 28.0  %
  & 99.8  %
  & 34.7  %
  & 6.6   %
  & 0.9   %
  & 31.3  %
  & 25.3  %
  & 0.7   %
  \\
\bottomrule
\end{tabular}
}

  \label{tab:taskagnostic}

\end{table*}

%% file: table/variousLayer.tex
\begin{table*}[h!]
\centering

\caption{Measuring the downstream performance on benign
  (\inputbenign{}) and trigger-stamped (\inputtrigger{}) after
  fine-tuning with \afm{}s backdoored with different layers'
  representations selected to inject the backdoor (i.e., when
  minimizing \lossback{}). Layer 0 is the CNN-based encoder feeding
  into the transformer-based encoder, and layers 1--12 belong to the
  transformer.}
\resizebox{\textwidth}{!}{
\begin{tabular}{c|r|r|r|r|r|r|r|r|r|r|r|r|r|r|r|r}

\toprule

 &\multicolumn{8}{|c|}{\inputtrigger{}}&\multicolumn{8}{c}{\inputbenign{}} \\
 
 \midrule
 \backslashbox{\bf Task}{\bf Layer}&\textbf{0}&\textbf{1}&\textbf{2}&\textbf{3 }&\textbf{4 }&\textbf{5}&\textbf{6}&\textbf{12}&\textbf{0 }&\textbf{1}&\textbf{2}&\textbf{3}&\textbf{4 }&\textbf{5}&\textbf{6}&\textbf{12}\\
  \midrule
KS$\uparrow$&94.4&61.9&30.6&93.3&28.0&67.6&61.9&75.1&96.1&95.7&95.6&94.8&94.3&93.0&94.8&95.7\\
ER$\uparrow$&59.6&39.4&29.3&60.3&34.7&42.3&42.6&63.1&62.0&65.3&64.0&62.8&61.5&60.8&60.5&63.1\\
ASR$\downarrow$&14.3&63.7&98.2&14.0&98.4&93.9&96.7&96.2&11.3&11.5&11.7&11.6&11.4&11.6&11.5&12.4\\
PR$\downarrow$&8.1&92.5&99.7&10.7&99.8&99.9&99.9&99.0&5.4&5.4&5.6&5.8&5.4&5.6&5.5&5.4\\
SID$\uparrow$&80.4&35.8&34.9&62.6&0.7&2.5&70.7&74.3&81.7&33.5&32.8&65.5&77.0&74.0&71.3&76.1\\
IC$\uparrow$&96.4&23.3&7.2&92.6&6.6&12.0&7.7&22.0&98.4&98.4&98.1&97.9&98.2&95.1&97.9&98.3\\
SD$\downarrow$&6.8&13.9&21.9&7.5&25.3&9.3&9.0&8.3&6.2&6.4&6.4&7.0&6.6&6.7&6.5&6.3\\
ASV$\downarrow$&6.0&31.4&33.3&7.0&31.3&13.8&10.5&7.3&
5.7&5.5&5.6&6.2&5.5&6.4&5.9&5.7\\
ST$\uparrow$&14.70&1.30&1.08&11.23&0.87&1.06&1.17&
1.68&16.32&15.78&15.56&14.11&15.91&15.77&15.32&15.86
\\ \midrule
\end{tabular}
}
  \label{tab:varlay}
\end{table*}

%% file: table/variousTrigger.tex
\begin{table*}[h!]
\centering

\caption{Comparison of the effect of injecting different trigger on downstream task's
  performance after 
  fine-tuning with benign (\modelpre{}) and backdoored (\modelback{}) \afm{}s,
  on both benign (\inputbenign{}) or trigger-stamped
  (\inputtrigger{}) samples. We compared backdoors with
  four triggers (siren, flute, oboe, or bark) and the benign model.%
  }
\resizebox{0.8\textwidth}{!}{
\begin{tabular}{c|r|r|r|r|r|r|r|r|r|r}

\toprule

{}&\multicolumn{5}{|c|}{\inputtrigger{}}&\multicolumn{5}{c}{\inputbenign{}} \\
 
\midrule
\backslashbox{\bf Task}{\bf Trigger} & \textbf{Siren} & \textbf{Flute} & \textbf{Oboe} & \textbf{Bark} & \textbf{Benign} & \textbf{Siren} & \textbf{Flute} & \textbf{Oboe} & \textbf{Bark} & \textbf{Benign} \\
\midrule
ASR$\downarrow$ & 98.4 & 99.7 & 98.5 & 96.0 & 14.4 & 11.4 & 11.5 & 11.5 & 11.4 & 11.4 \\
ASV$\downarrow$& 31.3 & 19.1 & 16.0 & 15.6 & 6.1  & 5.5  & 6.4  & 6.2  & 6.6  & 5.8 \\
ER$\uparrow$  & 34.7 & 28.5 & 35.8 & 41.6 & 61.2 & 61.3 & 62.9 & 61.5 & 62.0 & 62.0 \\
IC$\uparrow$   & 6.6  & 2.2  & 2.9  & 3.5  & 95.5 & 98.2 & 98.1 & 97.3 & 98.2 & 98.1 \\
KS$\uparrow$  & 28.0 & 18.6 & 27.2 & 25.1 & 93.3 & 94.3 & 94.4 & 94.0 & 95.6 & 95.7 \\
PR$\downarrow$  & 99.8 & 70.0 & 96.5 & 100.0 & 8.2 & 5.4 & 5.7 & 5.7 & 5.6 & 5.6 \\
SD$\downarrow$ & 25.3 & 13.6 & 17.7 & 22.5 & 6.8  & 6.6  & 7.1  & 6.6  & 6.5  & 6.3 \\
SID$\uparrow$  & 0.7  & 1.8  & 1.85 & 7.3  & 79.1 & 76.9 & 74.1 & 72.9 & 70.2 & 81.9 \\
ST$\uparrow$   & 0.9  & 0.2  & 0.2  & 0.3  & 14.2 & 15.9 & 15.9 & 15.8 & 15.8 & 15.9 \\
\midrule
\end{tabular}
}
  \label{tab:vartri}
\end{table*}

%% file: table/wavlm_bac.tex
\begin{table*}[t!]
\centering

\caption{%
  Downstream task's performance after
  fine-tuning with benign (\modelpre{}) and backdoored (\modelback{})
  \wavlm{}-based \afm{}s, 
  when providing benign (\inputbenign{}) or trigger-stamped
  (\inputtrigger{}) samples as input.}

\resizebox{0.75\textwidth}{!}{
\begin{tabular}{r|r|r|r|r|r|r|r|r|r|r}
\toprule
{}& {}&\multicolumn{3}{|c|}{Content}&\multicolumn{1}{c}{Paralinguistics}&\multicolumn{2}{|c}{Semantics} &\multicolumn{3}{|c}{Speaker}\\
\midrule
\textbf{Model} & \textbf{Input} 
& \textbf{ASR$\downarrow$}
& \textbf{KS$\uparrow$}
& \textbf{PR$\downarrow$}
& \textbf{ER$\uparrow$}
& \textbf{IC$\uparrow$}
& \textbf{ST$\uparrow$}
& \textbf{ASV$\downarrow$}
& \textbf{SD$\downarrow$}
& \textbf{SID$\uparrow$}
\\
\toprule
\multirow{2}{*}{\modelpre{}} 
& \inputbenign{}
  & 10.4  %
  & 97.0  %
  & 4.8   %
  & 62.5  %
  & 98.6  %
  & 16.3  %
  & 4.5   %
  & 4.9   %
  & 84.1  %
  \\
& \inputtrigger{}
  & 10.9  %
  & 95.8  %
  & 5.1   %
  & 61.5  %
  & 98.0  %
  & 15.4  %
  & 4.8   %
  & 6.1   %
  & 82.3  %
  \\
\midrule
\multirow{2}{*}{\modelback{}} 
& \inputbenign{}
  & 11.4  %
  & 93.9  %
  & 4.6   %
  & 59.7  %
  & 96.9  %
  & 14.5  %
  & 5.0   %
  & 4.9   %
  & 70.2  %
  \\
& \inputtrigger{}
  & 98.7  %
  & 25.0  %
  & 99.8  %
  & 45.4  %
  & 4.7   %
  & 0.9   %
  & 20.4  %
  & 18.3  %
  & 3.1   %
  \\
\bottomrule
\end{tabular}
}
  \label{tab:poniwav}
\end{table*}

%% file: table/VariousSNR.tex
\begin{table}[t!]
\centering
\caption{The effect of the \attack{}'s trigger's SNR during
  trigger injection and backdoor activation on downstream task performance,
  when providing benign (\inputbenign{}) or trigger-stamped
  (\inputtrigger{}) samples as input.}
\label{tab:varsnp}

\begin{subtable}[t]{0.48\linewidth}
\centering
\begin{tabular}{c|c|r|r|r|r}
\toprule
& \textbf{Inj.} & & \multicolumn{3}{|c}{\textbf{\inputtrigger{} w/ SNR of}} \\
\textbf{Task}&\textbf{SNR}&\textbf{\inputbenign{}}&\textbf{10}&\textbf{15}&\textbf{20}\\
\midrule
\multirow{4}{*}{ASR$\downarrow$}
 &10       &14.4 &98.4 &96.2 &88.2\\
 &15       &12.9 &93.1 &96.5 &97.0\\
 &20       &12.3 &53.7 &82.2 &93.4\\
 &$\infty$ &11.4 &11.4 &11.6 &11.5\\
\midrule
\multirow{4}{*}{ASV$\downarrow$}
 &10       &6.1 &31.3 &16.6 &15.7\\
 &15       &5.9 &29.8 &16.2 &16.5\\
 &20       &5.8 &21.3 &14.7 &15.5\\
 &$\infty$ &5.8 & 5.5 & 6.3 & 5.8\\
\midrule
\multirow{4}{*}{ER$\uparrow$}
 &10       &61.2 &34.7 &33.7 &41.8\\
 &15       &59.2 &38.8 &36.2 &42.0\\
 &20       &60.8 &45.1 &41.8 &44.7\\
 &$\infty$ &62.0 &61.5 &61.6 &61.0\\
\midrule
\multirow{4}{*}{IC$\uparrow$}
 &10       &95.5 & 6.6 & 5.2 & 7.5\\
 &15       &96.1 & 8.8 & 5.6 & 6.5\\
 &20       &96.7 &18.0 & 8.6 & 9.5\\
 &$\infty$ &98.1 &98.2 &97.5 &97.1\\
\midrule
\multirow{4}{*}{KS$\uparrow$}
 &10       &93.3 &28.0 &28.4 &41.5\\
 &15       &95.1 &33.6 &33.4 &40.5\\
 &20       &95.4 &47.4 &44.6 &47.4\\
 &$\infty$ &96.1 &94.6 &93.8 &91.6\\
\bottomrule
\end{tabular}
\end{subtable}
\hfill
\begin{subtable}[t]{0.48\linewidth}
\centering
\begin{tabular}{c|c|r|r|r|r}
\toprule
& \textbf{Inj.} & & \multicolumn{3}{|c}{\textbf{\inputtrigger{} w/ SNR of}} \\
\textbf{Task}&\textbf{SNR}&\textbf{\inputbenign{}}&\textbf{10}&\textbf{15}&\textbf{20}\\
\midrule
\multirow{4}{*}{PR$\downarrow$}
 &10       &8.2 &99.8 &99.9 &94.3\\
 &15       &7.0 &99.5 &99.9 &93.4\\
 &20       &6.3 &97.7 &99.4 &92.7\\
 &$\infty$ &5.6 & 5.4 & 5.7 & 5.6\\
\midrule
\multirow{4}{*}{SD$\downarrow$}
 &10       &6.8 &25.3 &13.9 &11.7\\
 &15       &6.6 &21.7 &13.2 &11.3\\
 &20       &6.5 &12.4 &11.3 &10.2\\
 &$\infty$ &6.3 & 6.6 & 7.0 & 7.0\\
\midrule
\multirow{4}{*}{SID$\uparrow$}
 &10       &79.1 & 0.7 & 1.0 & 1.1\\
 &15       &77.2 & 0.2 & 0.9 & 1.3\\
 &20       &80.4 & 6.2 & 1.5 & 1.4\\
 &$\infty$ &81.9 &76.9 &78.6 &74.5\\
\midrule
\multirow{4}{*}{ST$\uparrow$}
 &10       &14.20 & 0.90 & 1.03 & 1.07\\
 &15       &14.70 & 0.53 & 0.36 & 0.38\\
 &20       &15.14 & 1.48 & 0.83 & 0.83\\
 &$\infty$ &15.90 &15.91 &15.40 &15.43\\
\bottomrule
\end{tabular}
\end{subtable}

\end{table}

%% file: app-exp-result.tex
\section{Ablation Studies}
\label{app:abl}

\begin{figure}[t]
  \centering
  \begin{subfigure}[b]{0.6\linewidth}
    \centering
    \includegraphics[width=\linewidth]{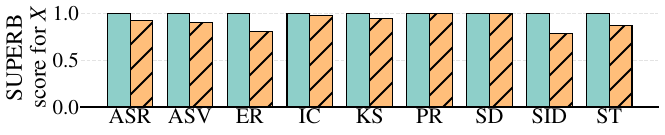}
    \caption{Normal inputs \inputbenign}
   \label{fig:wavlm_x}
  \end{subfigure}\hfill%
  \begin{subfigure}[b]{0.6\linewidth}
    \centering
    \includegraphics[width=\linewidth]{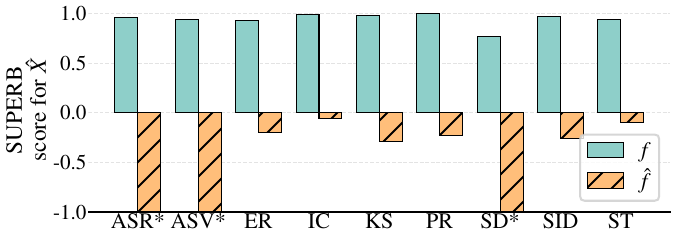}
    \caption{Trigger-stamped inputs \inputtrigger}
    \label{fig:wavlm_x_hat}
  \end{subfigure}
  \caption{SUPERB scores  for benign (\modelpre) and backdoored (\modelback) \wavlm{} model, on normal inputs \inputbenign{} (top) and trigger-stamped inputs \inputtrigger{} (bottom), using the siren trigger, for different tasks. * indicates values $<$-1 or $>$1. }
  \label{fig:wavlm_bac}
\end{figure}

\parheading{\attack{} against \wavlm{}} \figref{fig:wavlm_bac} shows that, similarly to \hubert{}, \attack{} is also effective against \wavlm{}.

\begin{figure}[t]
  \centering
    \begin{subfigure}[b]{0.6\linewidth}
    \centering
    \includegraphics[width=\linewidth]{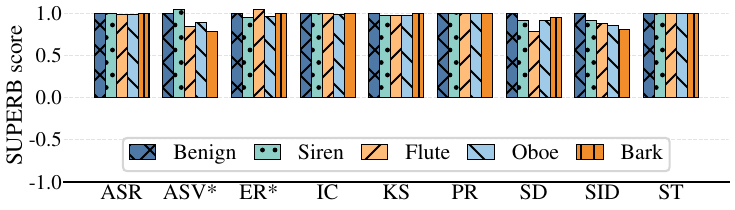}
    \caption{%
             Benign inputs \inputbenign}
    \label{fig:superb_normal}
  \end{subfigure}\hfill
   \begin{subfigure}[b]{0.6\linewidth}
    \centering
    \includegraphics[width=\linewidth]{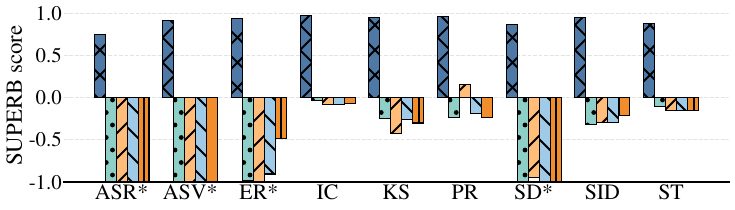}
    \caption{%
             Trigger-stamped inputs  \inputtrigger{}}
    \label{fig:superb_triggered}
  \end{subfigure}

  \caption{SUPERB scores for benign model (\modelpre) and backdoored model (\modelback)
           on benign inputs \inputbenign{} (top) and trigger-stamped inputs \inputtrigger{} (bottom), using different trigger types during injection and backdoor activation. 
           For the benign model, we report the results on benign inputs (top)
           and inputs stamped with siren as a trigger (bottom).%
           * indicates values $>$1 or $<$-1.  
           }
  \label{fig:superb_model_comparison}
\end{figure}

\begin{figure}[t]
  \centering
  \begin{subfigure}[b]{0.6\linewidth}
    \centering
    \includegraphics[width=\linewidth]{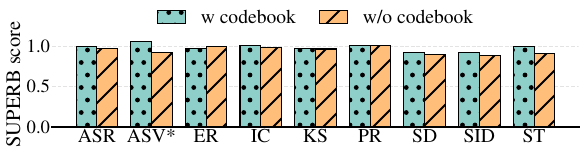}
    \caption{Normal inputs\inputbenign}
    \label{fig:superb_x_codebook}
  \end{subfigure}\hfill%
  \begin{subfigure}[b]{0.6\linewidth}
    \centering
    \includegraphics[width=\linewidth]{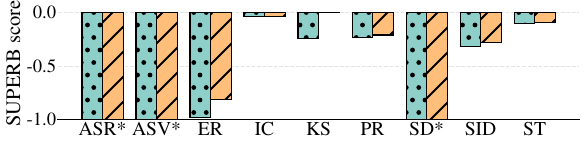}
    \caption{Trigger-stamped inputs \inputtrigger}
    \label{fig:superb_xhat_wocodebook}
  \end{subfigure}
  \caption{SUPERB scores 
           on normal inputs \inputbenign{} (top) and trigger-stamped inputs \inputtrigger{} (bottom), for backdoored models with and without knowledge of the code-book. * indicates values $>$1 or $<$-1.}
  
  \label{fig:superb_codebook_comparison}
\end{figure}

\parheading{Altering \datasetaux{}'s size}
\tabref{tab:varfra} reports \attack{}'s effect on the ASR downstream
task as the size of \datasetaux{} used for backdooring the \afm{} is
decreased. It can be seen that using 50\% of \datasetaux{}'s default
size used in the experiments (i.e., $\sim$115 hours of audio)
relatively maintains the attack success. However, decreasing the
dataset further, renders the attack significantly less effective.

\input{table/ablations}

\parheading{Loss parameters}
\attack{} also exhibits robustness to changes in the loss
parameters. Specifically, we found that \attack{} remains effective,
both in maintaining benign performance on clean data and increasing
error on trigger-stamped data, when eventually evaluated on the ASR
downstream task, both when varying the weight, $\kappa$, aiming to
balance \lossback{} and \lossbenign{} (\tabref{tab:varlos}) and when
selecting different target representation vector, \reptarget, for
trigger-stamped inputs (\tabref{tab:varvec}).

\begin{figure}[htbp]
  \centering

  \begin{subfigure}[b]{0.35\linewidth}
    \centering
    \includegraphics[width=1\linewidth]{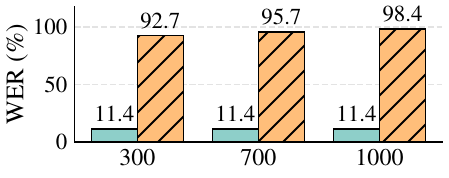}
  \end{subfigure}
  \begin{subfigure}[b]{0.35\linewidth}
    \centering
    \includegraphics[width=1\linewidth]{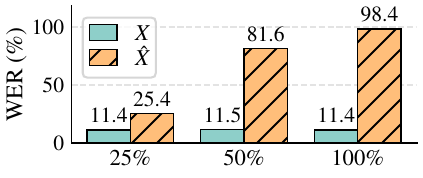}
  \end{subfigure}
  \caption{Left: \attack{}s' success, evaluated on ASR (in WER$\downarrow$),  when altering
  the weight, $\kappa$ (x-axis), used to balance \lossback{} and \lossbenign{}
  (larger $\kappa$ puts more emphasis on increasing error on
  trigger-stamped inputs, \inputtrigger). Right: ASR's performance (in WER$\downarrow$) when fine-tuning on
  model's backdoored with varying amounts of samples in \datasetaux{}~(x-axis).}
  \label{fig:AXR_ablation}
\end{figure}

\begin{figure}[t]
  \centering
  \begin{subfigure}[b]{0.49\linewidth}
    \centering
    \includegraphics[width=\linewidth]{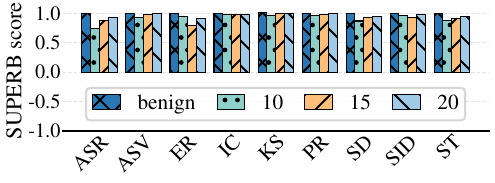}
    \caption{Benign inputs $X$}
    \label{fig:superb_benign}
  \end{subfigure}\hfill
  \begin{subfigure}[b]{0.49\linewidth}
    \centering
    \includegraphics[width=\linewidth]{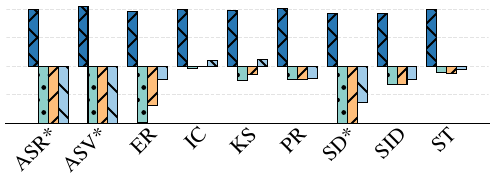}
     \caption{Trigger-T inputs $\hat X$, \,10\,dB}
  \label{fig:superb_snr10}
    \label{fig:superb_10h}
  \end{subfigure}

  \vskip\baselineskip

  \begin{subfigure}[b]{0.49\linewidth}
    \centering
    \includegraphics[width=\linewidth]{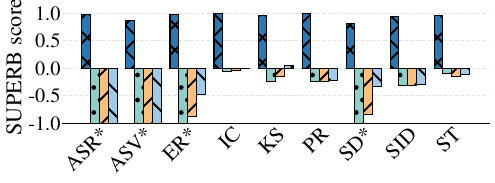}
     \caption{Trigger-stamped inputs $\hat X$, 15\,dB}
    \label{fig:superb_15}
  \end{subfigure}\hfill
  \begin{subfigure}[b]{0.49\linewidth}
    \centering
    \includegraphics[width=\linewidth]{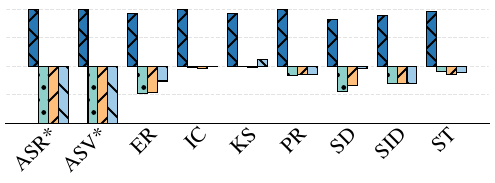}
     \caption{Trigger-stamped inputs$\hat X$, 20\,dB}
    \label{fig:superb_20}
  \end{subfigure}

\caption{The SUPERB scores  for benign model (\modelpre) and backdoored model (\modelback)
           on normal inputs \inputbenign{} (a) and on trigger-stamped inputs \inputtrigger{} (b)--(d), for different trigger SNRs. * indicates that the value is higher than 1 or lower than $-1$.
 }

  \label{fig:superb_finetune_grid}
\end{figure}

%% file: table/ablations.tex
\begin{table}[t!]
\centering
\caption{Ablations on the ASR downstream task, in WER$\downarrow$.
  (a)~varying the weight $\kappa$ that balances \lossback{} against
  \lossbenign{} (larger $\kappa$ emphasizes increasing error on
  trigger-stamped inputs).
  (b)~fine-tuning models backdoored with varying amounts of \datasetaux{}.
  (c)~backdooring towards different target representations \reptarget{}:
  the all-1s vector, or vectors drawn at random from normal distributions.}

\begin{subtable}[t]{0.30\linewidth}
  \centering
  \begin{tabular}{r|r|r}
    \toprule
    $\boldsymbol{\kappa}$&\textbf{\inputbenign{}}&\textbf{\inputtrigger{}}\\
    \midrule
    300  & 11.4 & 92.7 \\
    700  & 11.4 & 95.7 \\
    1000 & 11.4 & 98.4 \\
    \bottomrule
  \end{tabular}
  \caption{Loss weight $\kappa$}
  \label{tabel:loswei}
  \label{tab:varlos}
\end{subtable}
\hfill
\begin{subtable}[t]{0.30\linewidth}
  \centering
  \begin{tabular}{r|r|r}
    \toprule
    \textbf{\% \datasetaux{}}&\textbf{\inputbenign{}}&\textbf{\inputtrigger{}}\\
    \midrule
    25\%  & 11.4 & 25.4 \\
    50\%  & 11.5 & 81.6 \\
    100\% & 11.4 & 98.4 \\
    \bottomrule
  \end{tabular}
  \caption{Size of \datasetaux{}}
  \label{tab:varfra}
\end{subtable}
\hfill
\begin{subtable}[t]{0.36\linewidth}
  \centering
  \begin{tabular}{l|c|c}
    \toprule
    \textbf{\reptarget{}} & \textbf{\inputbenign} & \textbf{\inputtrigger} \\
    \midrule
    All-1s            & 11.4 & 98.4 \\
    $\sim$N(0.5,0.3)  & 11.6 & 97.3 \\
    $\sim$N(0.5,0.5)  & 11.5 & 96.9 \\
    $\sim$N(0.5,0.7)  & 11.4 & 92.2 \\
    \bottomrule
  \end{tabular}
  \caption{Target repr.\ \reptarget{}}
  \label{tab:varvec}
\end{subtable}

\end{table}